\documentclass[aps,reprint,twocolumns,pre,showpacs,superscriptaddress,floatfix,notitlepage,nofootinbib]{revtex4-1}
\usepackage{amssymb,amsmath,amsfonts} 
\usepackage{amsthm}
\usepackage{times}
\usepackage{eucal}
\usepackage{mathtools}
\usepackage{physics}
\usepackage{array}
\usepackage{graphicx,epsfig,xcolor}
\usepackage[colorlinks=true,citecolor=blue]{hyperref}
\usepackage{ulem}

\begin{document}

\title{Thouless Energy Challenges Thermalization on the Ergodic Side of the Many-Body Localization Transition}

\author{\'{A}ngel L. Corps}
   \email[]{angelo04@ucm.es}
    \affiliation{Departamento de Estructura de la Materia, F\'{i}sica T\'{e}rmica y Electr\'{o}nica, Universidad Complutense de Madrid, Av. Complutense s/n, E-28040 Madrid, Spain}

\author{Rafael A. Molina}
\email[]{rafael.molina@csic.es}
\affiliation{Instituto de Estructura de la Materia, IEM-CSIC, Serrano 123, E-28006 Madrid, Spain}
    
\author{Armando Rela\~{n}o}
\email[]{armando.relano@fis.ucm.es}
\affiliation{Departamento de Estructura de la Materia, F\'{i}sica T\'{e}rmica y Electr\'{o}nica \& GISC, Universidad Complutense de Madrid, Av. Complutense s/n, E-28040 Madrid, Spain}

\date{\today} 

\begin{abstract}

We study the ergodic side of the many-body localization transition in its standard model, the disordered Heisenberg quantum spin chain.  We show that the Thouless energy, extracted from long-range spectral statistics and the power-spectrum of the full momentum distribution fluctuations, is not large enough to guarantee thermalization. 
 We find that both estimates coincide and behave non-monotonically, exhibiting a strong peak at an intermediate value of the disorder.  
  Furthermore, we show that non-thermalizing initial conditions occur well within the ergodic phase with larger probability than expected. Finally, we propose a mechanism, driven by the Thouless energy and the presence of anomalous events, for the transition to the localized phase.

\end{abstract}

\maketitle

\section{Introduction}\label{introduction}
In classical mechanics there is a strong link between ergodicity and thermalization. However, the situation is different in quantum mechanics. The conjecture by Bohigas, Giannoni and Schmit (BGS) \cite{bgs} establishes that the spectral fluctuations of quantum systems with an ergodic classical analogue exactly follow random matrix theory (RMT) \cite{rmt,Haake2010}. Hence, deviations from ergodicity are usually identified via spectral statistics.

Alternatively, quantum thermalization is justified by the eigenstate thermalization hypothesis (ETH) \cite{Nandkishore2015,rigol2016,Tasaki1998,Rigol2008,Reimann2015,Reimann2018,Deutsch2018}, which refers to the properties of expected values of physical observables in the eigenstates of the Hamiltonian. It is normally accepted that a quantum system thermalizes if the diagonal fluctuations of these expected values decrease fast enough with the system size \cite{rigol2016}. This statement is believed to be a consequence of quantum chaos and, ultimately, of RMT. Notwithstanding, RMT is more strict regarding the behavior of these diagonal fluctuations: they must constitute an uncorrelated random sequence, a request only present in the most rigurous view of ETH \cite{erh}. Hence, the link between quantum chaos and thermalization is by no means established yet, as a complete connection between these two theories is still missing. Unfortunately, simulating large quantum systems is not feasible, so it is not clear whether ergodicity in the sense of BGS is mandatory for thermalization, or a less rigurous definition for the term is sufficient. 

In this Article we deal with this issue by investigating one of the most striking exceptions to thermal behavior in many-body quantum systems: the transition to many-body localization (MBL) \cite{Altman2018, Basko2006,Nandkishore2015,Alet2018,Abanin2019,Schreiber2015,Lukin2019,Choi2016,Smith2016,Roushan2017,Xu2018,Suntajs2019,Sierant2019,Altshuler1997,Luitz2015,Facoetti2016,Pino2016,Luitz2017,Mace2019,Filippone2016,Abanin2019,Sierant2019PRB,Lev2016,Suntajs2020}. The MBL phase is an insulating quantum phase of matter that emerges in some
disordered interacting many-body systems, like the paradigmatic one-dimensional
spin chain, when the disorder is large enough
\cite{Basko2006,Nandkishore2015,Alet2018,Abanin2019}. 
Several experiments in one-dimensional lattice fermions and bosons \cite{Schreiber2015,Lukin2019}, two-dimensional interacting bosons \cite{Choi2016}, trapped ultracold ions \cite{Smith2016}, and superconducting qubits \cite{Roushan2017,Xu2018} have found its signatures. Notwithstanding, its relevance in the thermodynamic limit (TL) is still under active discussion \cite{Suntajs2019,Sierant2019}. 
Its counterpart, the ergodic phase, in
which thermalization is normally expected, is usually not under such scrutiny. 
However, the transition between the MBL phase and the ergodic phase is not well understood yet. Griffiths effects where anomalously different disorder regions dominate the behavior are supposed to be very relevant close to the transition \cite{Agarwal2015,Gopalakrishnan2016}, although recent studies deny that relevance \cite{Weiner2019}.  The possibility of the existence of a non-ergodic but extended phase (a so-called \textit{bad metal}) between the ergodic and the MBL phases has been proposed but whether it survives in the TL is in doubt \cite{Altshuler1997,Luitz2015,Facoetti2016,Pino2016}. Numerical studies of the ergodic phase in spin models showing transitions to a MBL phase for large disorder have found that the ergodic phase shows subdiffusive dynamics and other non-trivial behavior \cite{Luitz2017,Mace2019,Filippone2016}. 
Yet, it is not clear if these properties are generic or system dependent and they are probably very much affected by intrinsic limitations of the numerical size scaling due to the exponential growth of Hilbert space dimensions. Overall, MBL and the associated transition continue to receive a great deal of attention from different viewpoints and approaches \cite{Kjall,Devakul,Luitz2016,Yu2016,Modak2015,Serbyn2015,Serbyn2017,Kokalj,Ros2015,Vosk2015,Potter2015}. 

Here, we deal with the disordered Heisenberg chain, where a transition from ergodic to MBL phases is expected to occur. We focus on the deviations from RMT that happen within the usually identified as ergodic region \cite{Altshuler1986,Altshuler1988,stockmann,Jensen1985,Srednicki1994,Deutsch1991}, and we study their consequences in ETH and thermalization. We report a neat connection between the Thouless energy,  $E_{\textrm{Th}}$ \cite{Edwards1972,Shapiro1993}, the energy scale beyond which spectral statistics deviate from RMT universal results, and the diagonal fluctuations of relevant observables. We find that these fluctuations cease to constitute an uncorrelated random signal beyond the scale set out by the Thouless energy, giving rise to another deviation from RMT. For small spin chains, this scale determines up to what extent the system thermalizes ---the smaller the Thouless energy, the more probable is to find a non-thermalizing initial condition. 
Furthermore, the transition from the more chaotic to the MBL region is triggered by an extended region in which the distribution of deviations from thermal equilibrium is very long-tailed, a region which is also characterized by a very small Thouless energy. This fact is compatible with other features associated to Griffith effects \cite{Griffiths1969,Luschen2017,Agarwal2015,Gopalakrishnan2016,Weiner2019, Luitz2016}. Finite-size scaling available to current computational capabilities suggests that this region does not shrink as the system size is increased, but its fate in the thermodynamic limit is still not clear. Hence, the following picture is compatible with our results: $(i)$ an integrable limit when the chain is disorder-free; $(ii)$ a narrow (almost) ergodic region, at small disorder; $(iii)$ an anomalous region, within the apparent ergodic phase, with a significant probability of finding non-thermalizing initial conditions; and  $(iv)$ the MBL phase, in which generic initial conditions are expected not to thermalize. 

The remainder of this paper is organized as follows. In Sec. \ref{thermalizationmechanism} we review the concept of quantum thermalization, governed by ETH, and comment on its connection with RMT. We argue that the deviations from RMT in many-body ergodic systems, identified by the Thouless energy, must have measurable consequences in the thermalization process. In Sec. \ref{secmodel} we introduce the model that we use: the Heisenberg chain, which has become the standard model to test many-body localization. Sec. \ref{thoulesseth} is devoted to a large part of our main results. We devise a novel approach to treat the diagonal fluctuations of the ETH so that they can be easily put in comparison with results that involve  long-range spectral statistics. We find that there exists a characteristic scale well-within the ergodic phase that can be identified in both measures. It behaves non-monotically and very approximately coincide for spectral statistics and the diagonal flucutations of observables. We then study thermalization by a quench protocol in this region, and find that it shows vast differences even within the ergodic phase, so deviations from RMT represent an important role, at least in finite systems. In Sec. \ref{secanomalous} we investigate the transition from the ergodic to the many-body localized phase. Our results suggest that the emerging structure is a precursor of the transition, heavily influenced by Griffiths effects. We find anomalously long-tailed distributions that do not show any scaling with the system size which, strictly, is incompatible with thermalization. Our results suggest that the ergodic region of the model is actually not as wide as put forward by some previous works. Finally, in Sec. \ref{secconclusions} we gather the main results of our work. 

\section{Thermalization and its mechanism}\label{thermalizationmechanism}

Let us consider an initial condition, $\ket{\psi(0)}$, evolving in an isolated quantum system with Hamiltonian $H$, $\ket{\psi(t)}=\exp\left(i H t/\hbar \right) \ket{\psi(0)}$. The key element to determine if this particular initial condition thermalizes is the behavior of long-time averages of expected values of physical
observables, 
\begin{equation}
\label{longtimeaverages}
\begin{split}
\langle \hat{O}\rangle_{t} &:= \lim_{\tau\rightarrow
  \infty}\frac{1}{\tau} \int_0^{\tau} \textrm{d}t \, \bra{\psi(t)}\hat{O}\ket{\psi(t)}  \\ &=\sum_n  \left| C_n \right|^2 \bra{E_{n}}\hat{O}\ket{E_{n}},
  \end{split}
  \end{equation}
where $\ket{E_n}$ represents the eigenstate with energy $E_n$, $H \ket{E_n} = E_n \ket{E_n}$, and $\left|C_n\right|^2 := \left|\left< \psi(0) \right| \left. E_n \right>\right|^2$ is the probability of finding the system in the eigenstate $\ket{E_n}$. For simplicity, we have assumed that the energy spectrum is not degenerate. 
  
Thermalization occurs if Eq. \eqref{longtimeaverages} is equal to the microcanonical average,
\begin{equation}\label{mea}
\langle \hat{O}\rangle_{\textrm{ME}}:=\frac{1}{\mathcal{N}}\sum_{E_n\in[E-\Delta E,E+\Delta E]}\bra{E_n}\hat{O}\ket{E_n},
\end{equation}
where $E$ is the (macroscopic) energy of the system, and $\Delta E$ a small energy window, $\Delta E/E \ll 1$, containing a large number of levels, ${\mathcal N} \gg 1$. 

It is well known that long-time averages like Eq. \eqref{longtimeaverages} remain close to an
equilibrium value under very generic circumstances
\cite{Reimann2008,Linden2009}, although some questions remain
open \cite{Hamazaki2018}. However, this equilibrium value is not
necessarily equal to Eq. \eqref{mea}. In classical mechanics, the link between the equivalent results is well supported by chaos. If the system is ergodic and mixing, any trajectory erratically explores the whole region of the phase space with energy $E$, and therefore long-time averages become equivalent to phase space averages restricted to the right value of the energy $E$ \cite{Pathria}.

In quantum mechanics, the situation is rather different. The equivalence between
microcanonical and long-time averages lies in the ETH
\cite{Jensen1985,Deutsch1991,Srednicki1994,Tasaki1998,Rigol2008,Reimann2015}.
In a few words, this theory states that a system is expected to thermalize for an observable $\hat{O}$ if the diagonal terms
$O_{nn}:=\bra{E_{n}}\hat{O}\ket{E_{n}}$, $n\in\{1,\ldots,N\}$ change with energy smoothly enough. 

To get a more detailed picture \cite{Deutsch2018}, let us consider that, regardless
of whether the system thermalizes or not, one has
\begin{equation}\label{diagonalterms}
   O_{nn}=\langle \hat{O}\rangle_{\textrm{ME}}+\Delta_{n},\,\,n\in\{1,\ldots,N\},
\end{equation} 
where $N$ denotes the size of the Hilbert space, and the quantity $\Delta_{n}$ represents how close the diagonal element $O_{nn}$ is to the microcanonical average, $\left< \cdot \right>_{\textrm{ME}}$; we
will call it \textit{ diagonal fluctuations}. The ETH requires that $\Delta_{n}$ decreases fast with the system size for thermalizing systems \cite{Deutsch2018}. In
its strong version, it demands that all the values of
$\Delta_{n}$ be negligible; for its weak version, it suffices
that most $\Delta_{n}$ fulfill this condition
\cite{Reimann2018,Hamazaki2018}.

These facts do establish a link between quantum thermalization and chaos, but not so strong as in classical mechanics.
 The quantum analogs of mixing classical systems give rise to energy spectra whose statistical properties coincide with those of RMT \cite{bgs}. Regarding the diagonal fluctuations, $\Delta_n$, RMT gives rise to an uncorrelated random signal with exponentially decaying width with the system size \cite{Deutsch2018}. This constitutes a stronger condition for $\Delta_n$ than the one demanded by the ETH, and it is only considered under certain circumstances \cite{erh}. Contrarily, $\Delta_{n}$ can show some
structure in integrable systems, due to the presence of additional quantum
numbers \cite{Peres1984,Lobez2016}.

Deviations from RMT are well known, even within regions identified as ergodic. In disordered many-body systems these deviations can be identified via the so-called Thouless energy \cite{Edwards1972,Shapiro1993}. In noninteracting disordered metals, this is an energy scale related to the typical time that a particle takes to diffuse across the sample. However, for interacting systems, the meaning of this quantity is still under active discussion, although there is some convincing evidence that it might be related to a complex anomalous diffusion process \cite{Bertrand2016}. As expected in the noninteracting limit, level statistics of interacting systems were shown to be well described by RMT universal results for eigenvalues separated by less than this quantity, but deviate towards the typical behavior for integrable systems at larger scales \cite{Bertrand2016}. However, the dynamical consequences of this fact are not clear at all. The emergence of a finite Thouless energy has been argued to be connected with Griffiths effects and the subdiffusive phase appearing on the ergodic phase, but this is still a subject that deserves further investigation as some questions remain open to this day. At the same time, ergodicity has been assumed if the Thouless energy grows fast-enough with the system size \cite{Suntajs2019,Suntajs2020}, but no stringent test regarding thermalization has been done to support this claim.  The main aim of this paper is to study the role played by this energy scale in the thermalization process, a topic in which, we believe, there has been little to no research.

\section{Model}\label{secmodel}
We work with the standard model for
MBL: a one-dimensional chain with two-body nearest-neighbor
couplings, $L$ sites, and onsite magnetic fields, the Heisenberg model \cite{Serbyn16,Bertrand2016,Torres2017,lfsantos,Buijsman2019},
\begin{equation}
  \label{model}
  \begin{split}
  \mathcal{H}&=\sum_{\ell=1}^{L}\omega_{\ell}\hat{S}_{\ell}^{z}\\&+J\sum_{\ell=1}^{L-1}\left(\hat{S}_{\ell}^{x}\hat{S}_{\ell+1}^{x}+\hat{S}_{\ell}^{y}\hat{S}_{\ell+1}^{y}+\lambda
  \hat{S}_{\ell}^{z}\hat{S}_{\ell+1}^{z}\right),
\end{split}\end{equation}
where $\hat{S}_{\ell}^{x,y,z}$ are the total spin operators at site $\ell\in\{1,\ldots,L\}$.
We choose $J=1$, $\hbar:= 1$. Periodic boundary conditions are applied, which minimize finite-size effects. For our simulations, we let $\lambda$ vary to study quenched dynamics, while we fix $\lambda=1$ to analyze eigenlevel statistics. Disorder is implemented by the
uniformly, randomly distributed magnetic fields $\omega_{\ell}\in[-\omega,\omega]$. For $\omega=0$, the chain is disorder-free and it gives rise to fully integrable dynamics that can be described by means of the Bethe-ansatz \cite{Nandkishore2015,rigol2016}. For intermediate values of $\omega$, the chain is believed to exhibit an ergodic phase where most initial conditions are expected to thermalize. The spectral statistics of this region are complex: they show a behavior close to the Gaussian orthogonal ensemble (GOE), following RMT, but with long-range deviations due to the Thouless energy. Overall, this metallic region is by no means a common one as a number of anomalous phenomena have been previously diagnosed. Close to the transition, a Griffith-like phase \cite{Lev2014,Lev2015,Luitz2015,Luitz2016,Agarwal2015,Znidaric2016,Ros2015,Gopalakrishnan2016} is responsible for slow subdiffusion and sublinear power-law growth of the entanglement entropy. For $\omega$ larger than a critical value which depends on the system size $L$, the model enters the many-body localized phase where (generic) initial conditions do not thermalize at all. Both the Bethe-ansatz and the MBL phases show Poissonian  spectral statistics, which means energy levels are here essentially uncorrelated. 

As commonly done in the literature, in this work we consider the the eigenvalues associated to the eigenstates of $\hat{S}^{z}:=\sum_{i}\hat{S}_{i}^{z}$. Since this operator commutes with the Hamiltonian, $[\mathcal{H},\hat{S}^{z}]=0$, we restrict ourselves to the sector $S^{z}=0$, where $S^{z}$ is the eigenvalue of the operator $\hat{S}^{z}$. Thus the dimension of the Hilbert space is $d=\binom{L}{L/2}$, but we will only consider the central $N=\binom{L}{L/2}/3$ eigenstates $\{\ket{E_{n}}\}_{n=1}^{N}$ to avoid border effects. Main results are shown for $L=16$, which gives $N=4290$, but for finite-size scaling considerations some results will also be shown for other values of $L$. 

\section{ Thouless energy and Eigenstate Thermalization Hypothesis}\label{thoulesseth} 

\subsection{Short-range spectral statistics}
By far, the most common indicator in the literature to identify the $\omega$ range for which systems show ergodic or integrable dynamics is short-range spectral statistics \cite{stockmann}. To this end, the distribution of the ratio of two-level spacings \cite{ratios,Oganesyan2007}, $P(r)$, has been employed to a great extent. This is just the distribution of $r$: the random variable taking on values 
\begin{equation}r_{n}:= \frac{E_{n+1}-E_{n}}{E_{n}-E_{n-1}},\,\,\,\,\forall n\in\{2,\ldots,N-1\}.\end{equation} 
  The set of energies is supposed to be in ascending order, $\{E_{1}\leq\ldots\leq E_{N}\}$. This spectral statistic captures short-range spectral correlations only and is not to be trusted when long-range correlations are to be investigated. Specifically, only the spectral properties of energies separated by level distances less than or equal to 2 can be described by this statistic. In particular, we make use of the equivalent measure $\langle \widetilde{r}\rangle$ where $\widetilde{r}$ is the random variable  \begin{equation}\label{minratios}\widetilde{r}_{n}:= \min\Bigg\{r_{n},\frac{1}{r_{n}}\Bigg\}\in[0,1],\,\,\,\forall n\in\{2,\ldots,N-1\}.\end{equation} The quantity $\langle\widetilde{r}\rangle$ is always defined, as opposed to $\langle r\rangle$, which diverges for a spectrum with Poissonian statistics. For the chaotic GOE, $\langle \widetilde{r}\rangle_{\textrm{GOE}}\approx 0.5307(1)$, while for the integrable (Poissonian) limit this is $\langle \widetilde{r}\rangle_{\textrm{P}}=2\ln 2-1$.

This information is complemented with the interpolating distribution of the ratios suggested in \cite{Corps2020}, \begin{equation}\label{distribution}
   P_{\gamma\beta}(r):= C_{\beta}\frac{(r+r^{2})^{\beta}}{\left[(1+r)^{2}-\gamma(\beta)r\right]^{1+3\beta/2}}.
\end{equation}  The generalized Dyson index $\beta\in[0,1]$ indicates the degree of chaos: $\beta=0$ is for Poisson whereas $\beta=1$ is for GOE. The ansatz for $\gamma=\gamma(\beta)$ in the Poisson-GOE transition proposed in \cite{Corps2020} has been used, and $\beta$ is obtained by fitting \eqref{distribution} to the numerical histograms for each value of the disorder parameter, $\omega$. Finally, $C_{\beta}$ are normalization constants implicitly verifying $\int _{0}^{\infty}\textrm{d}r\,P_{\gamma\beta}(r)=1$.

\begin{figure}[h]
\begin{tabular}{c}
\hspace{-1.4cm}\includegraphics[width=0.36\textwidth]{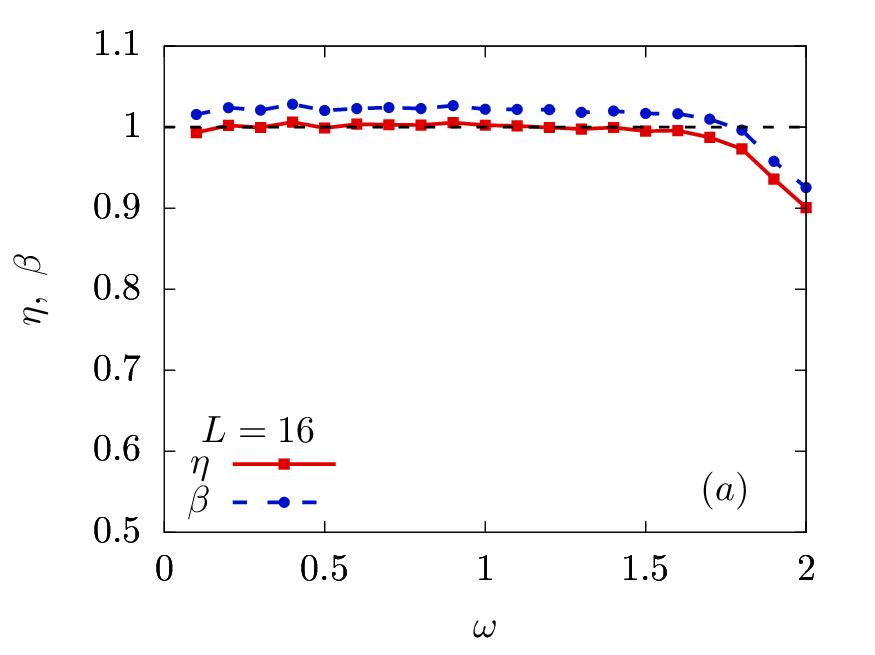} \\
\includegraphics[width=0.36\textwidth]{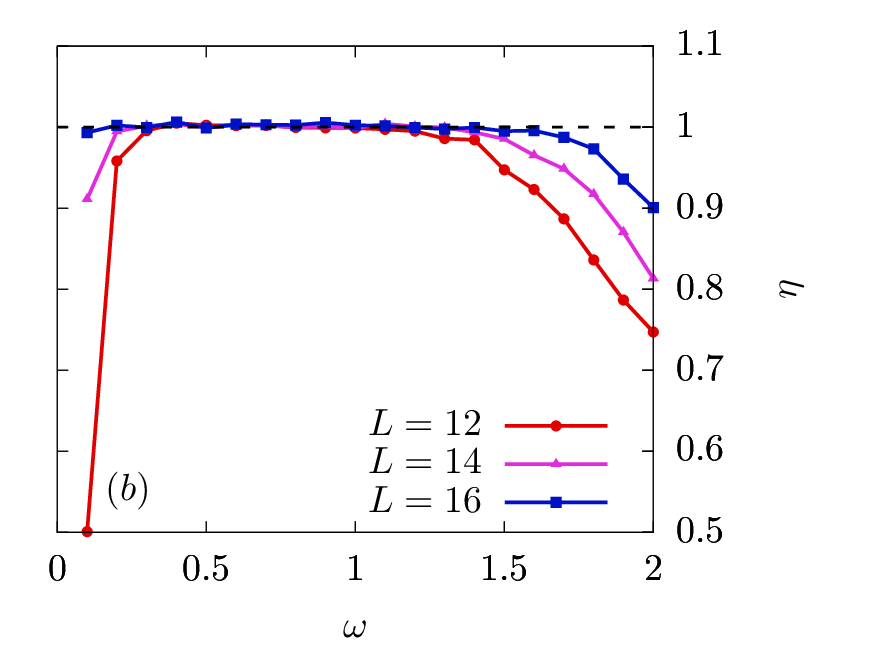}
\end{tabular}
\caption{Panel $(a)$: $\eta$ and Dyson index $\beta$ for $L=16$ as a function of the disorder parameter $\omega$. Panel $(b)$: $\eta$ for $L\in\{12,14,16\}$ as a function of $\omega$.  }
\label{ratios}
\end{figure}

To easily compare both estimates, the results for $\langle \widetilde{r}\rangle$ are shown after the rescaling $
\eta:=\frac{\langle\widetilde{r}\rangle_{\textrm{P}}-\langle\widetilde{r}\rangle}{\langle \widetilde{r}\rangle_{\textrm{P}}-\langle\widetilde{r}\rangle_{\textrm{GOE}}}$,
which is such that $\eta=0$ for Poisson and $\eta=1$ for GOE. These quantities are shown in Fig. \ref{ratios}. For all three values of $L$, there is a strong plateau that shrinks as $L$ is decreased. Conversely, short-range measures provided by the ratios suggest that as $L$ is increased the plateau that determines the size of the ergodic region should stretch in the directions of both large and small $\omega$. This is the region usually identified as ergodic in the literature \cite{Serbyn16,Sierant2019PRB,Alet2018,Bertrand2016}. The transition to the many-body localized phase is initiated at a certain value of the disorder strength that strongly depends on the system size, $L$. For $L=16$, this ergodic region covers about $0.2\lesssim\omega\lesssim1.8$, although the exact boundaries have not been completely delimited. However, the ratios only afford information about the distribution of eigenlevels separated by \textit{small distances}, and cannot capture in any way long-range spectral correlations, i.e., the statistical properties of eigenlevels further apart. We will see in the next subsection that this leads to important consequences. 

\subsection{Time series approach to the diagonal fluctuations} 

The results of the previous section are not sensitive to the existence of the Thouless energy, $E_{\textrm{Th}}$. Its value is
commonly obtained from long-range spectral statistics. 
For disordered spin-chains, 
the number variance $\Sigma^2(L)$ \cite{Bertrand2016} and the spectral form-factor $\mathcal{K}(\tau)$
\cite{Suntajs2019, Sierant2019} have been used.
A simpler and convenient alternative is given by the
$\delta_n$ spectral statistic \cite{conjetura,demo}. 
It
measures the distance between the $n$-th unfolded energy level,
 a dimensionless quantity obtained from the smooth part of the
cumulative level density $\varepsilon_n := \overline{N}(E_n)$, being
$E_n$ the $n$-th energy level \cite{misleadingsign}, and its
average value in an equiespaced spectrum, $\langle\varepsilon_{n}\rangle=n$, i.e.,
\begin{equation}\label{delta}
\delta_n := \varepsilon_n - n,\,\,\,\,\,n\in\{1,\ldots,N\}.
\end{equation}

Formally, $\delta_{n}$ can be seen as a time series signal where the discrete time is represented by the level order index $n$. Its power spectrum, $\langle P_{k}^{\delta}\rangle$, was shown to provide a neat characterization of fully chaotic and integrable systems in terms of the power-law decay $\langle P_{k}^{\delta}\rangle\simeq 1/k^{\alpha}$, where the exponent depends on level correlations and takes the value $\alpha=2$ for uncorrelated (i.e., integrable) spectra and $\alpha=1$ for quantum chaotic systems \cite{conjetura,Pachon2018,Gomez2005}, without making explicit reference to any random matrix ensemble. As mentioned above, to calculate $\delta_{n}$ knowledge of the cumulative spectral function is required. This function essentially gives the number of levels with energy less than or equal to a certain energy value $E$, and can be written $N(E)=\sum_{n=1}^{N}\Theta(E-E_{n})$, where $\Theta$ is the Heaviside step function. This function can be split into a smooth part $\overline{N}$ and a fluctuating part $\widetilde{N}$, i.e., $N(E)=\overline{N}(E)+\widetilde{N}(E)$. Separating the smooth cumulative level function from the fluctuations and then mapping the original energies $\{E_{n}\}_{n=1}^{N}$ onto new, dimensionless ones $\{\varepsilon_{n}\}_{n=1}^{N}=\{\overline{N}(E_{n})\}_{n=1}^{N}$ is called \textit{unfolding procedure} \cite{misleadingsign}, and it sets the mean level density to unity. It is under these circumstances that RMT universal predictions hold \cite{rmt,stockmann}. Different methods can be used to this end. When there is no theoretical underlying statistical theory that provides $\overline{N}(E)$, as in this case, it must be obtained by numerically fitting an staircase function with a polynomial of a certain degree. Note that $\delta_{n}$ is a dimensionless quantity because on the unfolded scale the transformed energies are simply numbers without units. 

The similarities between Eqs. \eqref{diagonalterms} and \eqref{delta} suggest a remarkable \textit{link between spectral statistics
and the ETH}. We note that $\Delta_n$ has the physical dimensions of the observable to which it refers. Thus, to make it dimensionless, as $\delta_n$ is, we normalize  by the
standard deviation $\sigma_{\Delta_{n}}=\langle \Delta_{n}^{2}\rangle$,
\begin{equation}\label{bardelta}
\widetilde{\Delta}_{n}:= \frac{\Delta_{n}}{\sigma_{\Delta_{n}}}=\frac{O_{nn}}{\sigma_{\Delta_{n}}}-\frac{\langle \hat{O} \rangle_{\textrm{ME}}}{\sigma_{\Delta_{n}}},\,\,n\in\{1,\ldots,N\}.
\end{equation}
The first result of this section comes from analyzing $\delta_n$ and $\widetilde{\Delta}_n$ with the same tools. In short, we extend the time series analysis approach that was initially conceived for the $\delta_{n}$ statistic to the diagonal fluctuations, and treat them both equivalently. It is the formal similarity between the two quantities that compels us to carry on such a procedure: while $\delta_{n}$ represents the deviation of the $n$-th excited level with respect to its value in an equiespaced spectrum, $\widetilde{\Delta}_{n}$ is a measure of the (normalized) deviation of the $n$-th diagonal fluctuation of quantum observables with respect to its microcanonical equilibrium value. 

As observables $\hat{O}$, we choose the full momentum distribution on a one-dimensional lattice with lattice constant set to unity, i.e.,

\begin{equation}
\hat{n}_{q}:= \frac{1}{L} \sum_{m,n=1}^L  \text{e}^{2 \pi i (m-n) q/L}\hat{s}^+_m \hat{s}^-_n, \,\,q\in\{0,\ldots,L-1\},
\label{observable}
\end{equation}
where $\hbar:= 1$ and
$\hat{s}^{\pm}$ are the usual ladder spin operators. 

\begin{figure}[h]
\includegraphics[width=0.48\textwidth]{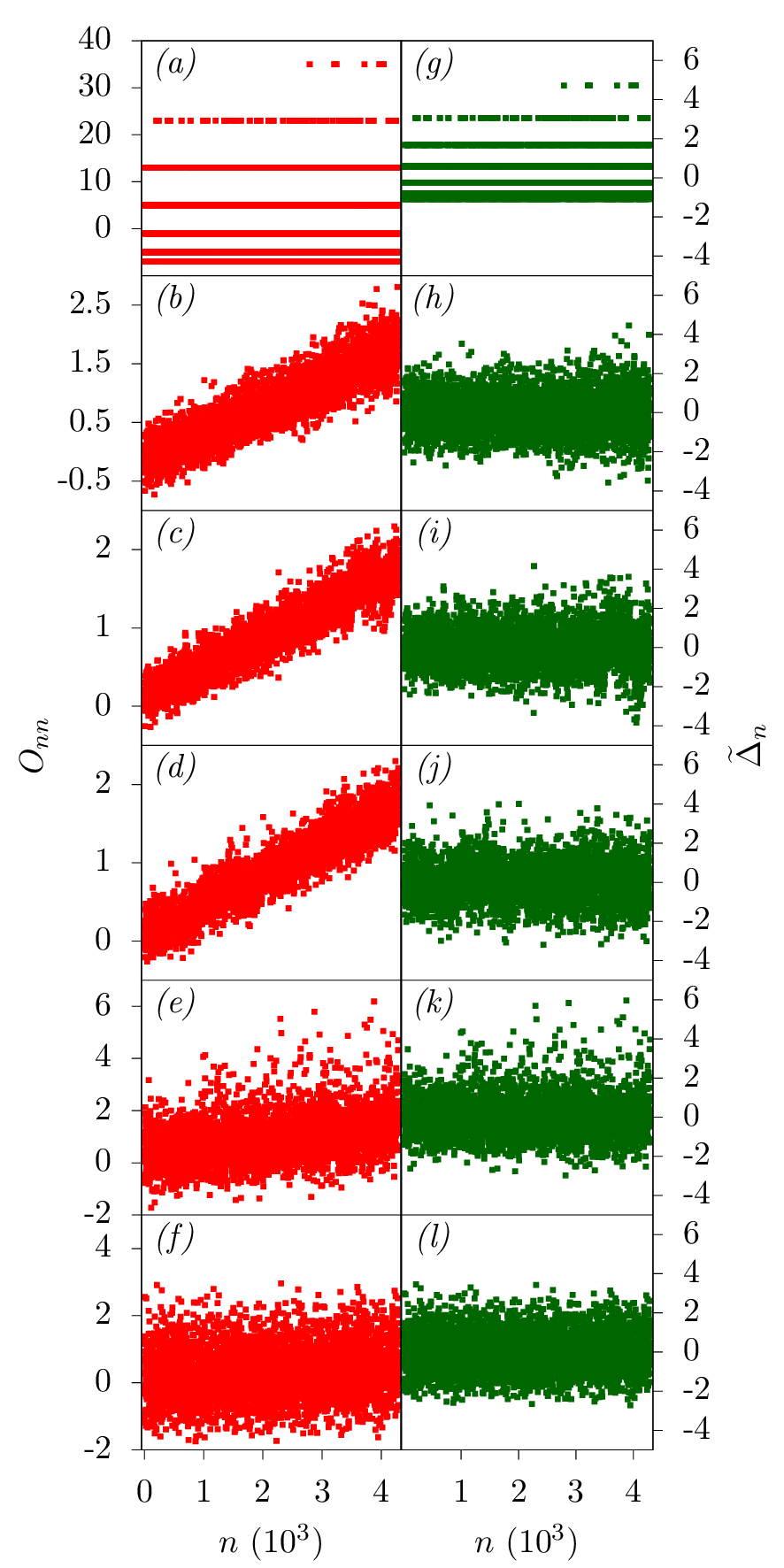}
\caption{Particular realization of the diagonal averages $O_{nn}$ (red symbols) and $\widetilde{\Delta}_n$ (green symbols) for the observable $\hat{n}_{0}$ as a function of $n\in\{1,\dots,4290\}$ for $L=16$. Panel $(a)$ and $(g)$ display $\omega=0$; panels $(b)$ and $(h)$, $\omega=0.6$; panels $(c)$ and $(i)$, $\omega=1$; panels $(d)$ and $(j)$, $\omega=1.4$; panels $(e)$ and $(k)$, $\omega=2.2$, and panels $(f)$ and $(l)$, $\omega=10$.}
\label{diagonal}
\end{figure}

In Fig. \ref{diagonal} we show a number of diagrams corresponding to a particular realization of the observable $\hat{n}_0$. Left panels display the raw diagonal averages $O_{nn}=\bra{E_{n}}\hat{n}_{0}\ket{E_{n}}$ for $\omega\in\{0,0.6,1,1.4,2.2,10\}$ (see caption for details). Right panels display $\widetilde{\Delta}_n$ for the same cases. The first remarkable fact is that both integrable limits, $\omega=0$ and $\omega=10$, behave in a very different way. $\hat{n}_0$ is constant of motion for $\omega=0$, and hence both $O_{nn}$ and $\widetilde{\Delta}_n$ show a clear structure of the kind of a Peres lattice, as expected for integrable dynamics \cite{Peres1984}. On the contrary, no such structure is seen in the MBL phase, represented in panels $(f)$ and $(l)$. Therefore, we expect observables $\hat{n}_q$ to show different features in the transition from ergodicity to both integrable limits. Besides this fact, it is difficult to extract conclusions from the rest of the panels, as they cannot be statistically told apart in an easy way and are very noisy. If we fix our attention in $\widetilde{\Delta}_n$, panels corresponding to $\omega=0.6$, $\omega=1$ and $\omega=10$ look very similar: the majority of the points are distributed over, roughly, $\widetilde{\Delta}_n \in (-2,2)$. Panel $(k)$, corresponding to $\omega=2.2$, shows a quite large set of extreme points outside this interval; and panel $(j)$, corresponding to $\omega=1.4$ seems to be in an intermediate situation. Notwithstanding, as these results constitute just one realization of the noise, no safe conclusions can be inferred. We will come back to this result later on. 

Besides the qualitative interpretation sketched above, relevant information can be obtained from analyzing the power spectrum $\langle P_{k}^{\widetilde{\Delta}}\rangle$ of $\widetilde{\Delta}_{n}$.  To obtain this quantity, the following steps need to be taken: \begin{itemize}
\item Fix the disorder strength $\omega$. 
    \item Fix $q\in\{0,\ldots,L-1\}$ and a particular realization of $\omega$. Let $W$ be the total number of realizations of each value of $\omega$. 
    \item Calculate the main trend of $O_{nn}$, Fig. \ref{diagonal}, which accounts for $\langle \hat{O}\rangle_{\textrm{ME}}$. Then, $\Delta_{n}=O_{nn}-\langle \hat{O}\rangle_{\textrm{ME}}$, $n\in\{1,\ldots,N\}$.  
    \item Find $\widetilde{\Delta}_{n}$ from its definition in Eq. \eqref{bardelta}.
    \item Apply discrete Fourier transform to $\widetilde{\Delta}_{n}$, \begin{equation}
       \mathcal{F}(\widetilde{\Delta}_{n}):= \frac{1}{\sqrt{N}}\sum_{n=1}^{N}\widetilde{\Delta}_{n}\exp\left(\frac{-2\pi ikn}{N}\right),
\end{equation}
     where $k\in\{1,2,\ldots,N-1\}$, and then take squared modulus, $P_{k}^{\widetilde{\Delta}}:=|\mathcal{F}(\widetilde{\Delta}_{n})|^{2}.$ This yields the power spectrum.
    \item Repeat for each of the $L$ values of $q$ and the $W$ values of $\omega$.
    \item Average over these $M=L\times W$ power spectra to obtain the mean estimator $\langle P_{k}^{\widetilde{\Delta}}\rangle$ in the usual way, \begin{equation}\langle P_{k}^{\widetilde{\Delta}}\rangle=\frac{1}{M}\sum_{i=1}^{M}\left(P_{k}^{\widetilde{\Delta}}\right)_{i},\,\,\,k\in\{1,\ldots,N-1\},\end{equation} where $\left(P_{k}^{\widetilde{\Delta}}\right)_{i}$ denotes the $i$th power spectrum. This yields the averaged power spectrum for the disorder value $\omega$ initially fixed. 
\end{itemize}

The microcanonical average is accounted for 
by fitting $O_{nn}$ to a polynomial of degree 4, which allows us to obtain the smooth part of diagonal terms as those in Fig. \ref{diagonal} by removing the fluctuations. As a consequence, the quantity $\Delta_{n}$ keeps track of the fluctuations of expected values of physical observables in the Hamiltonian eigenbasis instead, and it strongly oscillates around zero, i.e., its mean value, $\langle \Delta_{n}\rangle =0$ (see Fig. \ref{diagonal}). We remark that a fit to the main trend of $O_{nn}$ is replacing the actual $\langle\hat{O}\rangle_{\textrm{ME}}$, which would suffer from spurious effects that originate when averaging over a finite energy window \cite{misleadingsign}.  We average over
$W=40$ realizations of the magnetic field and 
$L=16$ observables given by Eq. \eqref{observable} in each
case, so $M=640$. As mentioned in Sec. \ref{secmodel}, we work with the $N=\binom{16}{8}/3=4290$ central eigenstates. 

\begin{figure}[h]
\begin{center}
\hspace{-1cm}
\includegraphics[width=0.45\textwidth]{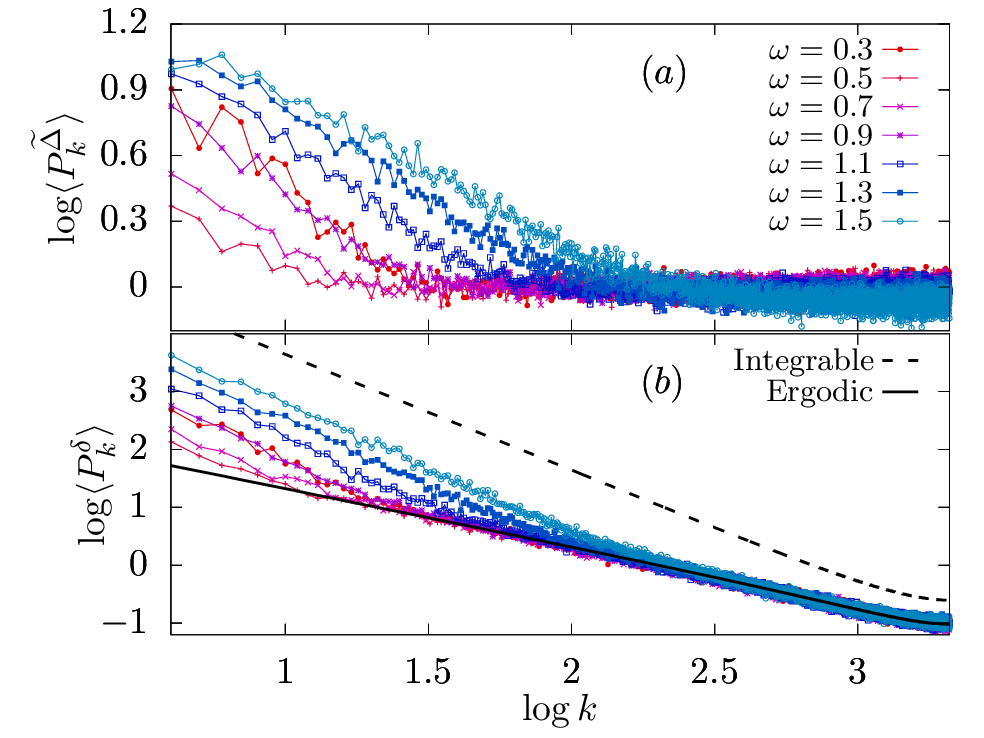}
\end{center}
\caption{ $(a)$: power-spectrum $\langle P_{k}^{\widetilde{\Delta}}\rangle$ of the diagonal fluctuations, Eq. \eqref{bardelta} as a function of $k\in\{1,\dots,N/2\}$. Results are averages over $640$ realizations of $N=4290$ expected values. $(b)$: power-spectrum $\langle P_{k}^{\delta}\rangle$ of the $\delta_{n}$ statistic, Eq. \eqref{delta}, for the system $\mathcal{H}(\lambda=1)$, Eq. \eqref{model}. Theoretical power-spectra for the ergodic, Eq. \eqref{powergoe}, and integrable cases \cite{demo}.  Results are averages over $100$ spectra of $N=4290$ levels each. The number of sites is always $L=16$ in this figure.}
\label{peth}
\end{figure}

Results in Fig. \ref{peth} focus on the power-spectra of the signals $\delta_{n}$ and $\widetilde{\Delta}_{n}$ on the expected ergodic region. For $L=16$, it covers $0.2\lesssim \omega \lesssim 1.8$ \cite{Sierant2019PRB,Serbyn16,lfsantos} (also see Fig. \ref{ratios}).
The power-spectrum of $\widetilde{\Delta}_{n}$, $\langle P_{k}^{\widetilde{\Delta}}\rangle:=\langle|\mathcal{F}(\widetilde{\Delta}_{n})|^{2}\rangle$, is shown in Fig. \ref{peth}$(a)$. We find that, for each $\omega$, large frequencies beyond a minimum value $k>k_{\min}$ exhibit, all, an equivalent weight, which corresponds to white, featureless (uncorrelated) noise. Conversely, small frequencies below such a minimum value $k<k_{\min}$ show an increasingly larger weight as $k$ is smaller, which immediately leads to a colored noise. This means that there is structure in the modes of the Fourier transform corresponding to large periods, i.e., to eigenstates very far apart from each other. Hence, $k_{\min}$ determines a characteristic scale. As RMT demands that diagonal fluctuations behave like an uncorrelated white noise, only Fourier modes shorter than the corresponding to $k_{\min}$ behave as expected for an ergodic region; larger scales show some structure. The lesser deviations from the ergodic expected value, zero, happen for intermediate values of the disorder close to $\omega_c=0.5$.

This behavior resembles that of $E_{\textrm{Th}}$ within the ergodic region and across the transition to MBL \cite{Suntajs2019,Sierant2019,Bertrand2016}. To delve into this coincidence, we invoke the $\delta_n$ statistic, Eq. \eqref{delta}. We follow the same procedure as before but now 
unfold with a polynomial of degree 6 to obtain the smooth cumulative level density and, therefore, the unfolded energies $\varepsilon_{n}=\overline{N}(E_{n})$, which is essentially the only requisite to calculate $\delta_{n}$. We have discarded 10\% of the energies closest to both spectrum edges before and after unfolding (that is, we discard a total of 20\% of the original levels), as these usually show anomalously large fluctuations and are not representative of the sample \cite{misleadingsign}.  Within the spectral form-factor approximation the reference GOE theoretical curve for this power-spectrum is given \cite{demo} by the free-parameter expression 
\begin{equation}\label{powergoe}
\begin{split}
    \langle P_{k}^{\delta}\rangle_{\textrm{GOE}}&=\frac{N^{2}}{4\pi^{2}}\left[\frac{\mathcal{K}\left(\frac{k}{N}\right)-1}{k^{2}}+\frac{\mathcal{K}\left(1-\frac{k}{N}\right)-1}{(N-k)^{2}}\right]\\& +\frac{1}{4\sin^{2}\left(\frac{\pi k}{N}\right)}-\frac{1}{12},
    \end{split}
\end{equation}
where $k\in\{1,2,\dots,N-1\}$, $N$ denotes the size of each spectrum in the ensemble over which the average has been performed, and $\mathcal{K}$ is the spectral form-factor \cite{rmt}, which for GOE can be written \begin{equation}
    \mathcal{K}(\tau)_{\textrm{GOE}}=\begin{cases}
{\displaystyle 2\tau-\tau\log(1+2\tau)},\,\,\,&\tau\leq1\\[0.2cm]
{\displaystyle 2-\tau\log\left(\frac{2\tau+1}{2\tau-1}\right)},\,\,\,&\tau\geq1
\end{cases}
\end{equation}

In Fig. \ref{peth}$(b)$ we show the power spectrum of $\delta_{n}$, \begin{equation}\langle P_{k}^{\delta}\rangle:=\langle |\mathcal{F}(\delta_{n})|^{2}\rangle=\left\langle\left|\frac{1}{\sqrt{N}}\sum_{n=1}^{N}\delta_{n}\exp\left(\frac{-2\pi i k n}{N}\right)\right|^{2} \right\rangle,\end{equation} $k\in\{1,2,\ldots,N-1\},$ calculated again by applying a discrete Fourier transform to the signal. It deviates from RMT universal behavior given in Eq. \eqref{powergoe} at roughly the same value as $\widetilde{\Delta}_{n}$, $k_{\textrm{min}}$. That is, \textit{ the characteristic scale given by the Thouless energy is directly transferred to the diagonal fluctuations}. Furthermore, this scale can be interpreted for the $\delta_n$ statistic in the same terms that for $\widetilde{\Delta}_n$: large frequencies beyond a minimum value $k>k_{\min}$ all lie on the theoretical GOE curve \eqref{powergoe}, i.e., eigenlevels separated by large distances corresponding to those frequencies $k$ follow universal chaotic behavior in the context of RMT. Conversely, small frequencies below a minimum value $k<k_{\min}$ show a clear deviation towards the integrable result for the power spectrum, indicating that the correlation between these energy levels only holds up to small energy distances associated to those frequencies $k$.

Remarkably, the transition from GOE to integrable statistics as the value of $\omega$ is increased within the ergodic phase is characterized by an increasing value of $k_{\min}$, but for sufficiently large values of $k$ the power spectrum always lies on the theoretical ergodic curve, regardless of $\omega$ (unless of course $\omega$ is large enough for the system to be well within the localized phase). \textit{ Exactly the same interpretation can be done with the diagonal fluctuations}. It is worth to point out that this is not the only possible shape for this kind of transition in terms of long-range spectral statistics. A number of systems show intermediate values for the power-law exponent, meaning that the power spectrum of $\delta_n$ statistic separates from the theoretical GOE curve at all frequencies, drifting towards the integrable result \cite{Gomez2005,Pachon2018,Santhanam2005,Relano2008}. In other systems, the deviation from RMT happens for high frequencies first \cite{Relano2008b}. As we have pointed out before, this means that the particular features of the transition from GOE to integrability in the $\delta_n$ statistics of the Heisenberg spin chain are directly transferred to the diagonal fluctuations. Both magnitudes preserve the typical properties of an ergodic system up to a characteristic scale, and deviate from the RMT behavior beyond this scale.

To interpret these results, we link the frequency $k$ to the scaled (dimensionless) time $\tau$ of the spectral form-factor, $\tau=k/N$ \cite{demo}. 
Then, we define a Thouless frequency, $k_{\textrm{Th}}:= N\tau_{\textrm{Th}}$, where $\tau_{\textrm{Th}}$ is the Thouless time. The Thouless time is the time scale associated to the Thouless energy. Thouless time and Thouless energy have been shown to be essentially inverse quantities not only in noninteracting but also in interacting systems (i.e., $E_{\textrm{Th}}\propto 1/\tau_{\textrm{Th}}$) \cite{Altshuler1986,Altshuler1988,Schiulaz2019}.  From it we obtain the inverse of the Thouless time $\ell_{\textrm{Th}}:= 1/\tau_{\textrm{Th}}=N/k_{\textrm{Th}}$, which represents a limiting scale for RMT-like behavior: energy levels within less than $\ell_{\textrm{Th}}$ are correlated like RMT spectra; those separated by more than $\ell_{\textrm{Th}}$ deviate from this behavior towards integrable-like correlations.  Note that this scale is dimensionless and represents how many energy levels are between two whose correlation is being calculated. A good estimate of the Thouless Energy is then $E_{\textrm{Th}}=\tau_{\textrm{Th}}^{-1}=\ell_{\textrm{Th}}/g(\epsilon)\propto \ell_{\textrm{Th}}$, where $\epsilon$ is the average energy, and $g(\epsilon)$, the density of states at such energy. As the frequency of Fig. \ref{peth}$(a)$ has the same physical meaning, a similar reasoning can be applied to this last case as well.

To determine $k_{\textrm{min}} \approx k_{\textrm{Th}}$ we choose the first $k$ for which the power-spectra fluctuate below ergodic expectations: the chaotic curve for $\langle P_{k}^{\delta}\rangle$ \cite{demo}, and zero for $\langle P_{k}^{\widetilde{\Delta}}\rangle$ (see discussion above in this section).
Next, we calculate the characteristic length $\ell_{\textrm{max}}:= N/k_{\textrm{min}}\approx\ell_{\textrm{Th}}$.
\begin{figure}[h]
\begin{center}
\includegraphics[width=0.46\textwidth]{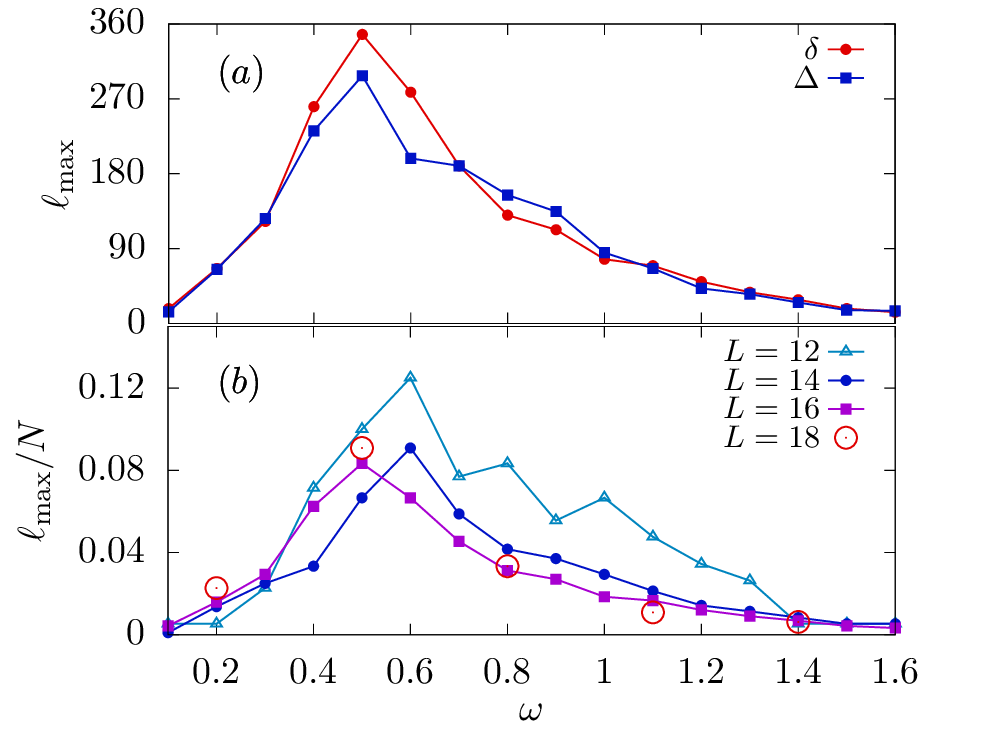}
\end{center}
\caption{ $(a)$: Value of the characteristic scale $\ell_{\textrm{max}}$ as a function of $\omega$ in the ergodic region for $\langle P_{k}^{\delta}\rangle$ and $\langle P_{k}^{\widetilde{\Delta}}\rangle$ and $L=16$. $(b)$: Normalized $\ell_{\textrm{max}}/N=k_{\textrm{min}}^{-1}$ as a function of $\omega$ for $\langle P_{k}^{\delta}\rangle$ and $L\in\{12,14,16,18\}$. }
\label{lmax}
\end{figure}
Results are shown in Fig. \ref{lmax}$(a)$.
We find good not only qualitative but also quantitative
agreement between these results for both power-spectra.
The value of $\ell_{\max}$ obtained from $\delta_{n}$, which measures the long-range statistic of eigenlevels, and $\widetilde{\Delta}_{n}$, which refers to the fluctuations of expected values of physical observables around the standard microcanonical average, are very approximately the same. This reinforces our previous conclusion: there exists a certain structure that manifests in both these two very different measures. Putting together Figs. \ref{diagonal} and \ref{lmax}, we can conjecture, for now, that the smaller $k_{\min}$, the more presence of anomalous points in $\widetilde{\Delta}_n$.

In Fig. \ref{lmax}$(b)$, we show $\ell_{\textrm{max}}/N$, representing such a \textit{ critical} scale. As $L$ is increased, fluctuations get gradually smoothed out, but the general structure remains very similar. 
This shows that the region $0.2 \lesssim \omega \lesssim 1.8$ is \textit{not ergodic in the sense of BGS}. It is characterized by a clear structure (usually associated with integrable-like, i.e., non-ergodic phenomena in the sense of BGS) which remains approximately unchanged irrespective of $L$: $\ell_{\max}/N$ has a maximum for $0.5\lesssim\omega\lesssim0.6$ (at which RMT behavior only holds within a scale equivalent to the $10 \%$ of the studied levels), and decreases for both larger and smaller values of disorder.  

These results need to be put in comparison with those afforded by the more common ratios (see Fig. \ref{ratios}). For the Heisenberg spin chain, as exemplified by the $\delta_{n}$ spectral statistic, short-range results can be quite misleading, as the Thouless energy behaves non-monotically and shows complex, strongly disorder-dependent behavior that is \textit{absolutely} absent from the homogeneous picture provided by the ratios. What the $\delta_{n}$ statistic tells us, as opposed to the ratios (or any other measure of short-range spectral statistics), is that even in the supposedly fully ergodic region there is an emergent structure that appears \textit{only between eigenlevels separated by more than a critical distance}, precisely $\ell_{\max}$, so these effects can in no way be diagnosed with measures that \textit{only afford information about eigenlevels separated by two levels or less}. Additionally, finite-size scalings have revealed in the past \cite{Santos2010,Torres-Herrera2014,Torres2017} that in the thermodynamic limit the influence of the Bethe-ansatz region \textit{may} shrink to a single point $\omega=0$ (i.e., an arbitrarily small $\omega>0$ would immediately take the system into the chaotic regime) and that the opposite side, the localized phase, could also be reduced to a single critical point, located at very high disorder.  However, Fig. \ref{lmax}$(b)$ shows no trace of such a scaling behavior: the `most ergodic' point seems to be fixed around $0.5\lesssim \omega\lesssim 0.6$, irrespective of $L$, i.e., there does not seem to be any clear scaling whatsoever. Thus, it does not seem very reasonable to expect the emerging structure in $\ell_{\max}$ to be a finite-size effect arising from the fact that there are two limiting integrable regions. Although we lack the evidence to state that this situation might be representative for very large values of $L$, we do have proof indicating that, at least for the finite values of $L$ usually considered, the structure in the characteristic scale is robust.

\subsection{Quenched dynamics on the expected ergodic phase}
From the previous results, the following question arises: do these (small) deviations from the RMT behavior entail measurable consequences in equilibrium states? In other words, how do these spectral properties manifest dynamically in terms of equilibration? The variance $\sigma_{\Delta_{n}}^{2}=\langle\Delta^2_n \rangle$  indicates whether typical states thermalize or not, and previous results suggest that it is small enough within the apparent ergodic region \cite{Nandkishore2015}. However, if one understands that ergodicity means that (almost) any initial condition thermalizes (as in the BGS result in RMT), this may still be insufficient.

We investigate here the probability of finding non-thermalizing initial conditions on the ergodic side. The analysis will be later extended to a greater region. We conjecture that \textit{ the key to this probability is the number of significantly populated eigenstates, $\mathcal{N}$}. If $\mathcal{N} < \ell_{\textrm{Th}}$, diagonal fluctuations behave as an uncorrelated white noise at all scales within the populated window, and no  anomalous effects may be expected. Otherwise, the emerging structure of diagonal fluctuations can impede thermalization for certain initial conditions: the larger the ratio $\mathcal{N}/\ell_{\textrm{Th}}$, the more likely to find a non-thermalizing one. The consequence is an \textit{anomalous region with a significantly large ratio of non-thermalizing initial conditions}.

\begin{figure}[h!]
\begin{center}
\includegraphics[width=0.46\textwidth]{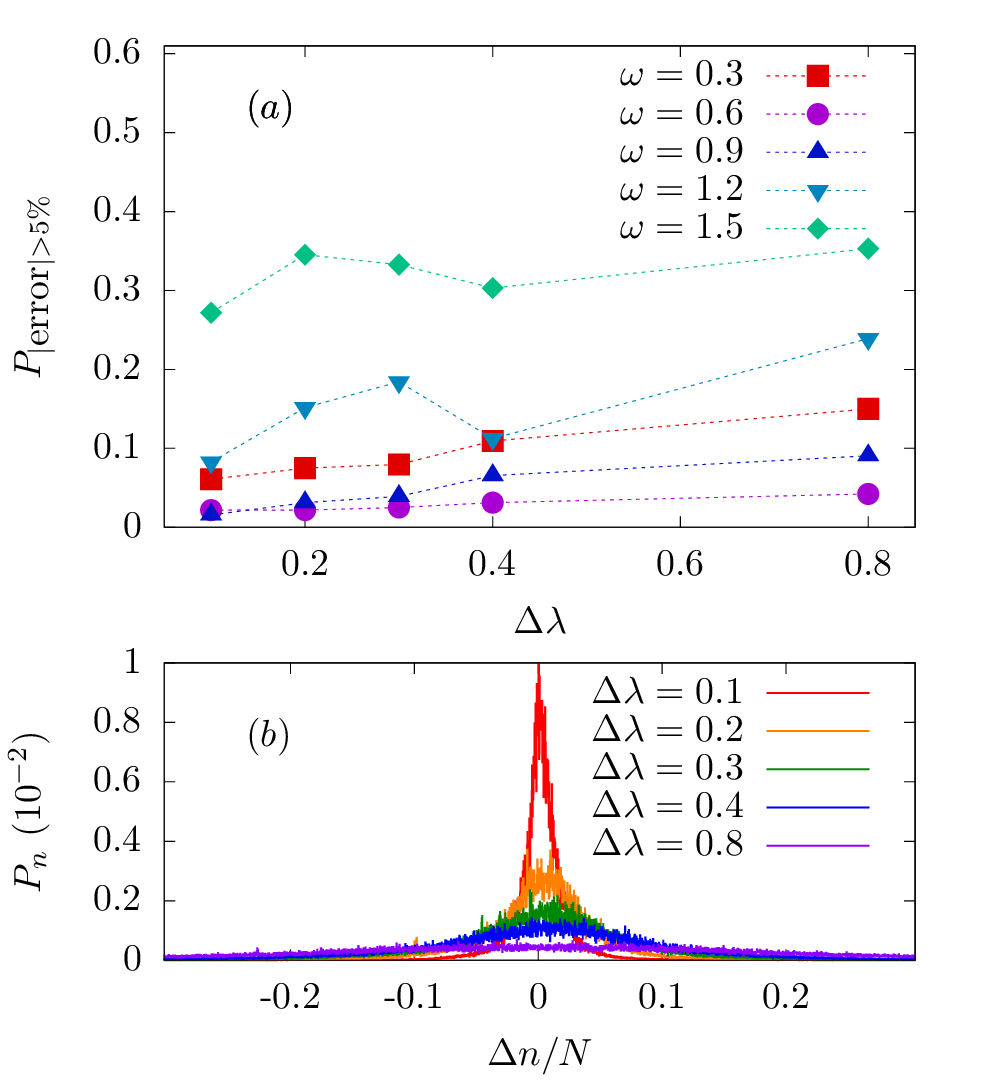}
\end{center}
\caption{$(a)$: probability that the relative difference between $\langle \hat{n}_{q}\rangle_{\textrm{ME}}$ and $\langle \hat{n}_{q}\rangle_{t}$ be greater than $5\%$ (see text for details).
The errorbars represent the standard deviation of the mean.  $(b)$: average of the eigenstate populations for the five quench sizes, obtained from $20$ disorder realizations for each $\omega$, and $\Delta \lambda\in\{0.1,0.2,0.3,0.4,0.8\}$. The $x$ axis represents the ratio of populated levels with respect to $N$, with $\Delta n=0$ being the centre of the populated window.  }

\label{termo}
\end{figure}

To test this conjecture,  we start from the central state of a certain initial value for $\lambda$ in Eq. \eqref{model}, $\lambda_{\textrm{i}} > 1$, and quench it onto $\lambda_\textrm{f}=1$,  with the same values for the random magnetic field, $\omega_n$. The size of this quench, $\Delta \lambda := |\lambda_\textrm{i} - \lambda_\textrm{f}|$, determines the width, $\mathcal{N}$, of the resulting state. We work with five different values of $\omega$ and five different quench sizes.

In Fig. \ref{termo}$(a)$ we plot the probability that the relative error of the difference between the long-time, $\left< \hat{n}_{q} \right>_t$, and the microcanonical, $\left< \hat{n}_{q} \right>_{\textrm{ME}}$ averages be greater than $5\%$, for different values of the disorder within the ergodic region, and different quench sizes. We have averaged over all $16$ observables and $40$ realizations of the random magnetic field. The microcanonical average is 
obtained with $41$ eigenstates around the expected energy of the initial state. We gather the following conclusions from this figure. First, the probability of finding a non-thermalizing initial condition increases as $\ell_{\textrm{max}}$ decreases. The case with $\omega=1.5$ is very significative. Although it is still within the ergodic region, the probability of finding an initial condition which deviates more than a 5$\%$ of the microcanonical prediction fluctuates around $30\%$. Second, this probability is generally higher for larger quenches, which populate larger number of levels.
In Fig. \ref{termo}$(b)$, a disorder average of the eigenstate populations for the five quench sizes is shown. We display the probability of finding the system in the eigenstate $\ket{E_n}$, $P_n$, versus the position of the eigenstate with respect to the central level, $\Delta n$ (that is, $\Delta n/N=0$ for the central level, and, for example, there are $n=0.1N$ levels between the one labelled with $\Delta n/N=0.1$ and the central one, where $N=4290$ is the total number of considered levels). This panel shows that the distribution of populated levels is quite wide in all the cases. Under many circumstances, the populated window resulting from the quench can be wider than $\ell_{\textrm{max}}$, and it is clearly seen that, the larger the quench, the wider the populated energy window. The case with the greatest $\ell_{\textrm{max}}$, $\omega=0.6$, is the least sensible to $\Delta\lambda$, and the one showing the smallest probability of such anomalous events. The cases $\omega\in\{0.3,0.9\}$ show a neat increase of this probability with $\Delta\lambda$. For $\omega\in\{1.2,1.5\}$, the behavior is more erratic, probably because the initial condition is wider than $\ell_{\textrm{max}}$ for all quench sizes.

\begin{figure}[h!]
\begin{center}
\includegraphics[width=0.44\textwidth]{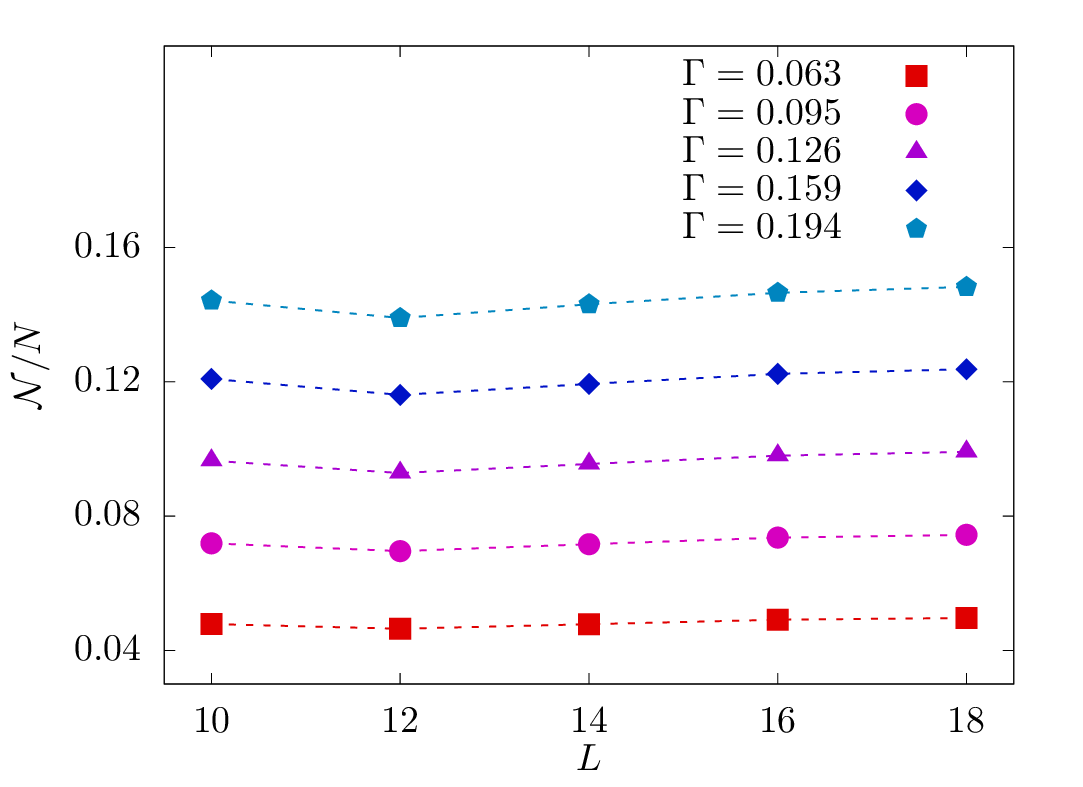}
\end{center}
\caption{$\mathcal{N}/N$ for  $\Gamma\in\{0.063,0.095,0.126,0.159,0.194\}$ from eigenlevels of Eq. \eqref{model} with $\omega=0.6$. Averages are taken over 1000 ($L=10,12,14$), 100 ($L=16$), and 20 ($L=18$) realizations. }

\label{termo2}
\end{figure}

These results are compatible with our previous statement. \textit{ The probability of non-thermalizing initial conditions decreases with increasing values of} $\ell_{\textrm{max}}$.

Fig. \ref{termo2} provides a hint about the consequences of this fact in the thermodynamic limit. We show how the number of populated eigenstates after a typical quench changes with $L$. We assume that the width of the corresponding initial state grows as expected for a canonical equilibrium state  \cite{Pathria}, $\sigma_{E}\propto\sqrt{L}$. Then, we study the number of levels, $\mathcal{N}$, populated in an energy window of width $\sigma_{E}=\Gamma\sqrt{L}$, being $\Gamma$ representative of the quench size. 
The ratio $\mathcal{N}/N$ remains approximately constant, suggesting that typical quenches may be wider than $E_{\textrm{Th}}$ even in the thermodynamic limit. We cannot extrapolate our results to much larger (macroscopic) systems, but they suffice to conclude that investigating quenched-dynamics requires a huge number of energy levels, a number that seems to grow linearly with the size of the system's Hilbert space, $\mathcal{N}\propto N$. Therefore, spectral statistics from a very small number of levels around the central one, like those afforded by, e.g., the shift-invert method \cite{shiftinvert}, are not enough to capture all the features associated to quenched dynamics. 

\section{Anomalous phenomena around the transition point}\label{secanomalous}

The picture emerging from these results is the following. The expected ergodic region has a subregion, $0.5 \lesssim \omega \lesssim 1$, in which the probability of non-thermalizing events is quite low. However, there exists another subregion, $1 \lesssim \omega \lesssim 1.8$, in which non-thermalizing events are high-probable enough to suspect that thermalization is not guaranteed for any initial condition. This is compatible with the much-debated Griffiths effects, appearing in some systems that display a transition to MBL from the ergodic side \cite{Griffiths1969,Luschen2017,Agarwal2015,Gopalakrishnan2016,Weiner2019}. 

To deepen into this matter, we have performed a stringent numerical test involving quenches of width $\Delta \lambda=0.4$ for different values of the system size $L\in\{10,12,14,16\}$. For the first case, $1000$ different realizations of the random magnetic field have been performed; for $L=12$, the number of realizations is $500$, and for $L=14$ and $L=16$, we have performed $240$ different realizations. As the corresponding total Hilbert space sizes are very different (from $d=252$ for $L=10$ to $d=12870$ for $L=16$), we have used different microcanonical windows: $\Delta E = 21$ for $L=10$; $\Delta E=31$ for $L=12$, and $\Delta E=41$ for $L=14$ and $L=16$. We have checked that small changes in these windows do not alter the results.  In this numerical experiment, we consider not only the ergodic phase of the chain but also the many-body localization edge, where the transition is supposed to take place. 

\begin{figure}[h!]
\hspace*{-0.4cm}
\begin{center}
\includegraphics[width=0.50\textwidth]{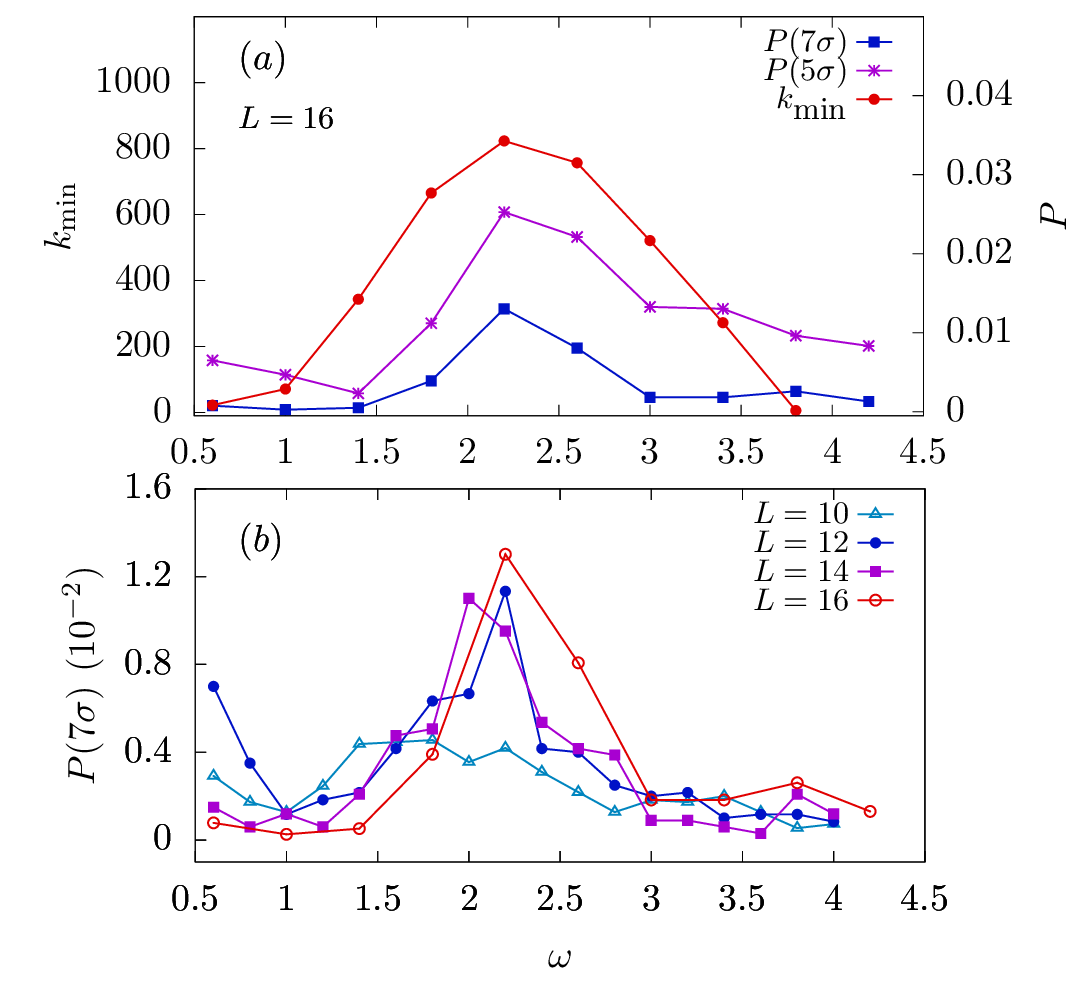}
\end{center}
\caption{$(a)$: Comparison of $k_{\min}$, obtained from $\langle P_{k}^{\widetilde{\Delta}}\rangle$, and $P(5\sigma)$ and $P(7\sigma)$ for $L=16$. $(b)$: $P(7\sigma)$ as a function of the disorder strength $\omega$ for $L\in\{10,12,14,16\}$.
Results correspond to expected values of the the full momentum distribution, $\hat{n}_{q}$, $q\in\{0,\ldots,L-1\}$ in the eigenstates of Eq. \eqref{model}. }
\label{fig4nueva}
\end{figure}

The link between these facts and the presence of correlations in $\widetilde{\Delta}_n$ is further explored in Fig. \ref{fig4nueva}. In Fig. \ref{fig4nueva}$(a)$ we show a comparison between the probabilities of anomalous events and the value of $k_{\min}$ obtained from the power spectrum of the diagonal fluctuations around the microcanonical average. The procedure to calculate $k_{\min}$ is the same as before. To obtain the probabilities of anomalous events, we have fitted a Gaussian distribution to the histograms of the relative differences between $\langle \hat{n}_{q}\rangle_{\textrm{ME}}$ and $\langle \hat{n}_{q}\rangle_{t}$, obtaining the standard deviation $\sigma$. From this result, we have calculated the probabilities of events whose relative deviation from $\langle \hat{n}_{q}\rangle_{\textrm{ME}}$ is larger than $5\sigma$ and $7\sigma$. For $\omega\approx 0.5$, the value of $k_{\min}$ is minimal, corresponding to very large values of $\ell_{\max}$. This means that the signal $\widetilde{\Delta}_{n}$ gives here basically an uncorrelated white noise. As $\omega$ is increased, the emerging structure extends to all scales throughout the entire range of disorder values, until it reaches it maximum at $\omega\approx 2.2$ (corresponding to minimal values of $\ell_{\max}$ instead). As $\omega$ is further increased into the depths of the localized phase, the structure in $k_{\min}$ gradually disappears, leaving behind quite a symmetric pattern around $\omega\approx 2.2$.  As the MBL phase is approached, $\widetilde{\Delta}_n$ turns back into an uncorrelated white noise (which gives minimal values of $k_{\min}$), but with much larger width, explaining why thermalization is not expected in this phase for any initial condition \cite{Nandkishore2015,Altman2018}. It is interesting to observe that these characteristic scales are also mimicked by the probability of anomalous events beyond $5\sigma$, $P(5\sigma)$ and $7\sigma$, $P(7\sigma)$, as obtained from a Gaussian distribution. This result shows that there exists quite a wide region, $1.5 \lesssim \omega \lesssim 3$, i.e. centered around the transition region from the ergodic to the MBL phase for $L=16$, with \textit{a very large probability of anomalous events}. Furthermore, these facts provide a quantitative explanation for the picture gathered from Fig. \ref{diagonal}. From panels $(h)$ to $(l)$ of that figure, we concluded that the probability of finding a large value for $|\widetilde{\Delta}_n|$ is larger for intermediate values of $\omega$, $\omega=2.2$, than for values representing both the more chaotic region, $\omega=0.6$, and the MBL phase, $\omega=10$. Results displayed in panel $(a)$ of Fig. \ref{fig4nueva} corroborate this idea.

The probability of anomalous events beyond $7\sigma$ as a function of the disorder for different values of $L$ is shown in Fig. \ref{fig4nueva}$(b)$. The probability of such events is similar for all values of $L$ (except for $L=10$, which may be not be representative as the size is too small) and, in any case, it shows no sign of a scaling behavior, i.e., these anomalies do not seem to decrease for larger chains. This leads us to the following conjecture, stating that this structure formation is a precursor of the transition: \textit{ the transition onto the MBL phase from the ergodic phase is initiated by an increase of the probability of anomalous, non-thermalizing, initial conditions, within the apparent ergodic region and driven by the Thouless energy}. This is directly linked to heavily long-tailed distributions near the MBL edge that have been diagnosed in the past as well \cite{Luitz2016} and several other anomalies characterizing the transition \cite{Devakul,Filippone2016,Luitz2015,Agarwal2015,Bertrand2016,Mace2019,Luschen2017}, notably including, but not limited to, manifestations of Griffiths effects and the presence of slow dynamics.

\begin{figure}[h!]

\begin{center}
\hspace*{-0.8cm}
\includegraphics[width=0.59\textwidth]{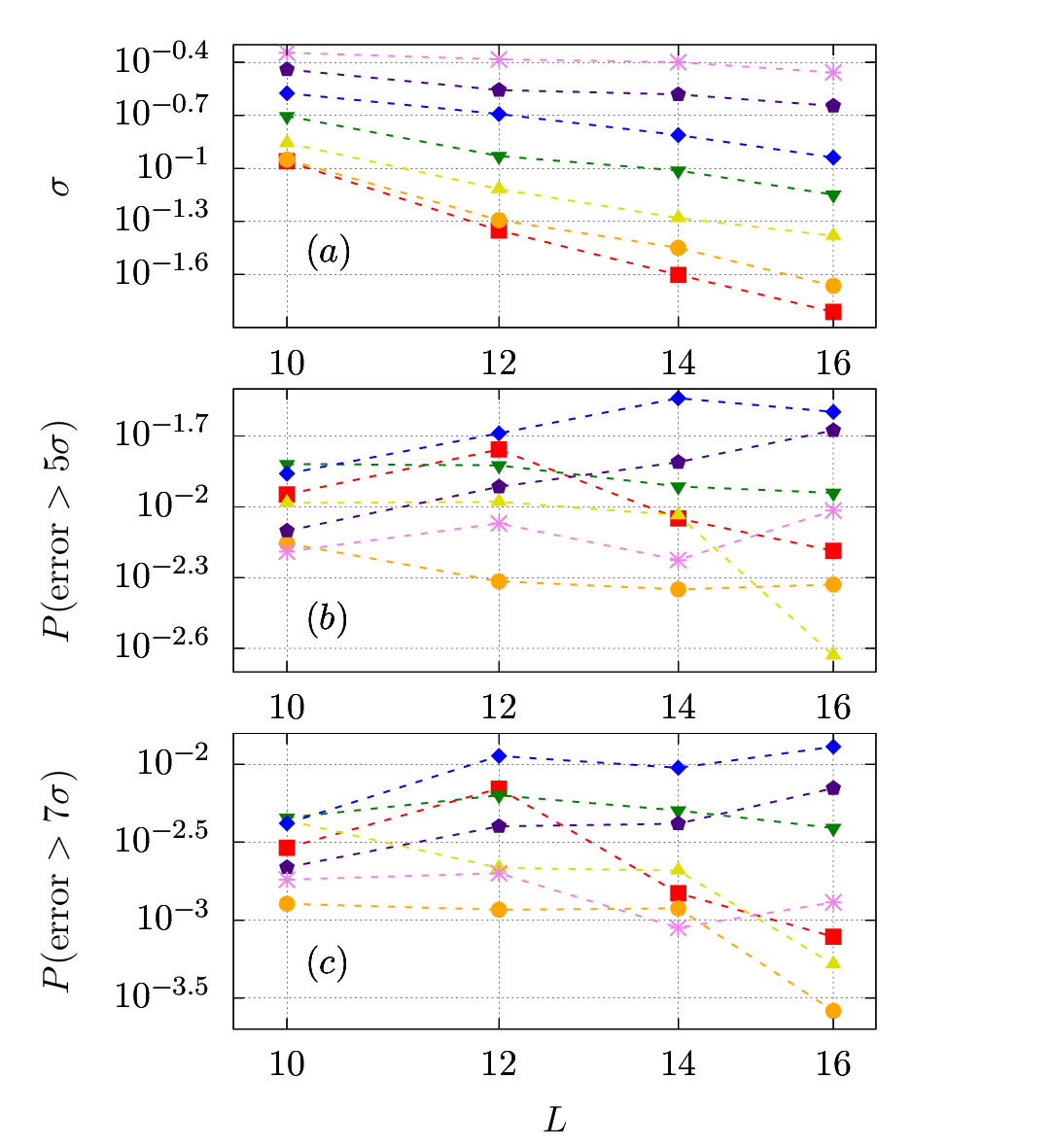}
\end{center}
\caption{$(a)$:  Scaling of the standard deviation $\sigma$ obtained from fitting the relative difference between time $\langle \hat{n}_{q}\rangle_{t}$ and microcanonical $\langle\hat{n}_{q}\rangle_{\textrm{ME}}$ averages to a Gaussian distribution as a function of the system size. $(b)-(c)$: Scaling of the probability of anomalous events beyond $5\sigma$, $P(5\sigma)$, and $7\sigma$, $P(7\sigma)$, as a function of the system size (log-log scale is used for all three panels). 
Results correspond to expected values of the the full momentum distribution, $\hat{n}_{q}$, $q\in\{0,\ldots,L-1\}$ in the eigenstates of Eq. \eqref{model}. Colors/point types represent disorder strengths $\omega\in\{0.6,1,1.4,1.8,2.2,2.6,3\}$ from bottom to top in panel $(a)$. Double logarithmic scale is used in all panels. }
\label{scalingsigma}
\end{figure}

The consequences in the thermalization process are explored in Fig. \ref{scalingsigma}. Fig. \ref{scalingsigma}$(a)$ shows a scaling of the value of $\sigma$ obtained as explained above. Even though it is quite daring to extrapolate these results to much larger (macroscopic) systems, the usual conclusion gathered from them is that they seem compatible with a fully ergodic region in the entire parameter space $0.6<\omega<3$ in the thermodynamic limit, as it displays a power-law decay when the system size is increased. Clearly the exponent of this power-law decay depends on the disorder strength as it is greater for $\omega\in\{0.6,1\}$, well-within the ergodic phase, than it is for very large values of $\omega$. Notwithstanding, it is recommendable to study more phenomena before accepting this conclusion. Fig. \ref{scalingsigma}$(b)-(c)$ shows a scaling of the probability of anomalous events $P(5\sigma)$ and $P(7\sigma)$ with the system size $L$. Here, we find two main results. First, the probability of such events is \textit{much larger than expected from a Gaussian distribution}, throughout the entire region $\omega\in[0.6,3]$, as the Gaussian probabilities of occurrence beyond $5\sigma$ and $7\sigma$ are, respectively, $P(5\sigma)_{\textrm{Gaussian}}\approx 5.7\cdot 10^{-7}$ and $P(7\sigma)_{\textrm{Gaussian}}\approx 2.6\cdot 10^{-12}$. These theoretical values are obviously a lot smaller than those obtained numerically for both cases, by several orders of magnitude. Second, the power-law decay that is present in panel $(a)$ is absent from panels $(b)$ and $(c)$. For large values of the system size a decrease of the probability of anomalous events can only be found for small values of $\omega$, in particular those closest to the point where the characteristic distance $\ell_{\max}$ is largest, whereas for others there is no evidence of such (the probability of anomalies is approximately the same for large $L$ or even increases). This is again a manifestation of very long-tailed distributions around the transition \cite{Luitz2016}. And it seems enough to call into question the na\"{i}ve extrapolation inferred from panel $(a)$ of Fig. \ref{scalingsigma}. Results shown in panels $(b)$ and $(c)$ of the same figure show that the analysis of thermalization is much more involved, since anomalous non-thermalizing events are highly probable before the MBL phase is reached, i.e., they are not rare at all. 

\begin{figure}[h!]
\begin{center}
\hspace*{-0.7cm}
\includegraphics[width=0.46\textwidth]{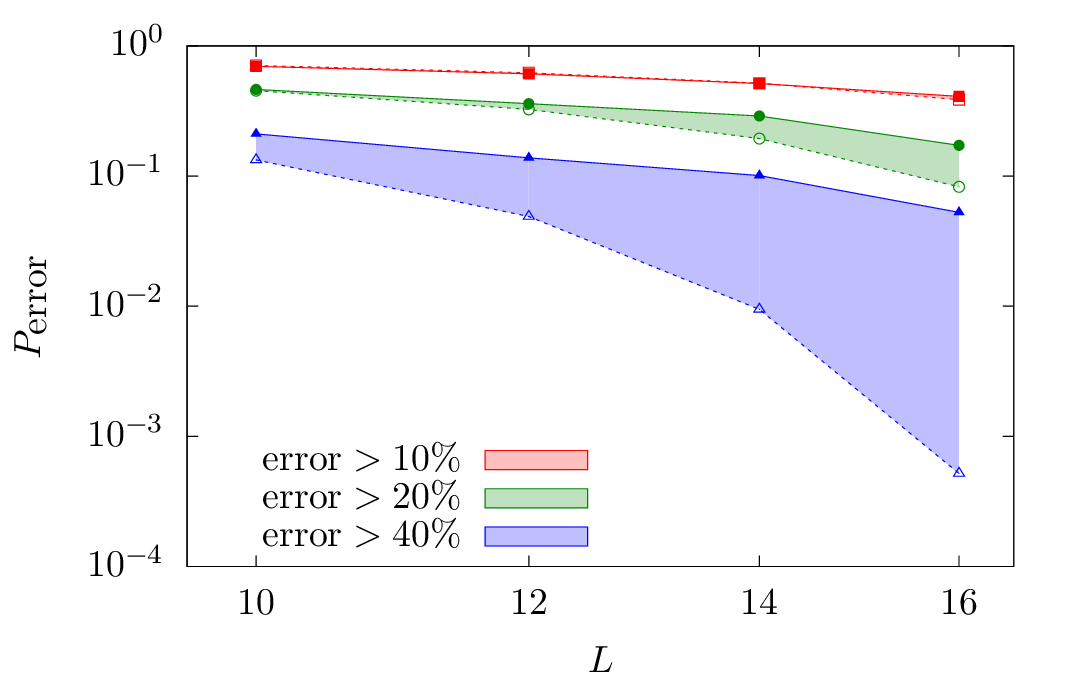}
\end{center}
\caption{ Scaling of the probability of initial conditions further than $10\%$, $20\%$ and $40\%$ from the equilibrium microcanonical average as a function of the system size for $\omega=2.2$. Solid lines with filled points represent the numerical value of $P_{\textrm{error}}$, while dotted lines with empty points show the corresponding value as obtained from a Gaussian distribution. 
Results correspond to expected values of the the full momentum distribution, $\hat{n}_{q}$, $q\in\{0,\ldots,L-1\}$ in the eigenstates of Eq. \eqref{model}. Double logarithmic scale is used. }
\label{scaling_w22}
\end{figure}

Finally, in Fig. \ref{scaling_w22} we plot a scaling of the probability of finding initial conditions further away than 10\%, 20\% and 40\% from its microcanonical equilibrium value in a fully ergodic region. The disorder value is here fixed at $\omega=2.2$, corresponding to the maximum of $k_{\min}$ and the probability of anomalous non-thermalizing events as seen in Fig. \ref{fig4nueva} and Fig. \ref{scalingsigma}. We compare here the expected result for a Gaussian distribution with the width, $\sigma$, obtained from Fig. \ref{scalingsigma}$(a)$ (dotted line and empty symbols), with the one obtained from numerics (solid line and full symbols); in all the cases, the shadowed region highlights the difference between the expectation and the numerical result. These results reinforce our previous conclusion. The probability of non-thermalizing events decrease much slower than expected from the Gaussian distributions underlying the ETH. And it is not clear at all how to extrapolate to very large (macroscopic) systems. This entails that, at least for finite systems, thermalization should in principle not be expected for every initial condition within a wide region covering part of the apparent ergodic phase. Further investigation is needed to even conjecture what happens in the thermodynamic limit.

\section{Conclusions}\label{secconclusions}

 We have provided a numerical, stringent study of the interplay between long-range spectral statistics and the diagonal fluctuations of physical observables around the microcanonical equilibrium value in the paradigmatic model for many-body localization, the Heisenberg spin-1/2 chain. There is a strong link between these two indicators of the dynamics of a quantum system. We have studied the power spectrum of the $\delta_{n}$ statistics to compare spectral correlations with the results coming from Random Matrix Theory, while analogously we have studied the power spectrum of the diagonal fluctuations of representative observables $\Delta_n$ to analyze the eigenstate thermalization hypothesis, which underlies the ability of a quantum system to thermalize.

 The power spectrum of these two quantities are quantitatively characterized by the same characteristic scale $\ell_{\max}$. In the case of spectral correlations, the $\delta_n$, the value of $\ell_{\max}$ indicates that Random Matrix Theory-like correlations between levels hold up to this scale. Dividing $\ell_{\max}$ by the density of states gives the famous Thouless energy scale, $E_{\textrm{Th}}$. In the case of $\Delta_{n}$, $\ell_{\max}$ indicates that the signal is essentially a white noise with little to no structure up to this scale, which is in turn associated with the Gaussian distribution of thermalizing events. Thus, the Thouless energy, the scale below which RMT universality breaks down, behaves non-monotically on the ergodic phase and its maximum is fixed constant at $0.5\lesssim\omega\lesssim0.6$ regardless of the system size (i.e., it does not scale). This puts under scrutiny the existence of a full ergodic region below a certain critical value of the disorder. Contrary to what is usually inferred from short-range spectral statistics, like the distribution of ratios of consecutive level spacings, in finite systems there exists no ergodic plateau between the integrable limit, $\omega=0$, and the MBL phase, but a region with a complex structure, with its most chaotic part around $\omega \sim 0.5$. 

 Then, we have presented numerical evidence implying that non-thermalizing events are significantly probable even within the apparent ergodic region of the Heisenberg chain, as a consequence of the previous structure. We have shown that ergodic RMT-like behavior only holds for quite short energy scales. The small value of the Thouless energy favours the existence of these anomalous initial conditions. The distribution of thermalizing events from the ETH acquires very long tails as we have shown by studying the probability of error between the microcanonical and time averages by a quench protocol. Studying these quantities as a function of disorder paints a very complex picture of the transition between the ergodic and the MBL phase. The minimum of $\ell_{\max}$ is found around $\omega\approx 2.2$ where we also find the largest probabilities of extreme values. At $\omega\gtrsim 3.6$ the system has completely entered the localized phase where the width of the distribution is much larger implying that generic initial conditions do not thermalize in this phase. These results give rise to the following picture. First, a more or less ergodic region, with small width for this distribution, and a small probability of anomalous events. Then, an intermediate extended phase, in which the width is still small, but the probability of anomalous events is largely increased. And finally, the MBL phase, characterized by a more Gaussian but wider distribution of relative deviations from thermal equilibrium. This complex picture does not change significantly when we increase the size of the system, although a lot of caution is needed when trying to extrapolate to larger sizes in spin models of this kind. Our results seem incompatible with a fully ergodic region emerging in the thermodynamic limit in the \textit{entire} parameter range $0<\omega\lesssim 3.6$ (for $L=16$). They also suggest that ergodicity in the sense of the BGS conjecture may be a necessary condition to guarantee thermalization for (almost) any initial condition. In this sense, studying short-range spectral correlations alone is not enough to understand chaos and thermalization in many-body quantum systems. It is important to fully grasp the complexity that emerges in the structure of long-range spectral statistics and the diagonal fluctuations of observables.  Another important conclusion is that a large number of energy levels and eigenstates are necessary to study the consequences of non-equilibrium dynamics and thermalization. Sudden quenches, the usual procedure to track the relaxation to equilibrium in small quantum systems, do significatively populate large number of energy levels. The results in this paper show that the structure of long-range spectral statistics and diagonal fluctuations of representative observables becomes highly complex within this range.

\acknowledgements
This work has been supported by the Spanish Grants
Nos. FIS2015-63770-P (MINECO/ FEDER) and PGC2018-094180-B-I00 (MCIU/AEI/FEDER, EU), CAM/FEDER  Project  No. S2018/TCS-4342 (QUITEMAD-CM) and CSIC Research Platform PTI-001.


\begin{thebibliography}{100}

\bibitem{bgs} O. Bohigas, M. J. Giannoni, and C. Schmit, \textit{Characterization of chaotic quantum spectra and universality of level fluctuation laws}, Phys. Rev. Lett. \textbf{52}, 1 (1984).

\bibitem{rmt} M. L. Mehta, \textit{Random matrices} (Academic Press, New York, 1991).

\bibitem{Haake2010} F. Haake, Quantum Signatures of Chaos (Springer,
Berlin, 2010).

\bibitem{Nandkishore2015} R. Nandkishore and D.A. Huse, \textit{Many-body localization and thermalization in quantum statistical mechanics}, Ann. Rev. Condens. Matter Phys. \textbf{6}, 15 (2015).

\bibitem{rigol2016} L. D'Alessio, Y. Kafri, A. Polkovnikov, and M. Rigol, \textit{From quantum chaos and eigenstate thermalization to statistical mechanics and thermodynamics}, Advances in Physics, 65:3, 239-362 (2016).

\bibitem{Tasaki1998} H. Tasaki, \textit{From quantum dynamics to the canonial distribution: general picture and a rigorous example}, Phys. Rev. Lett. \textbf{80}, 1373 (1998).

\bibitem{Rigol2008} M. Rigol, V. Dunjko, and M. Olshanii, \textit{Thermalization and its mechanism for generic isolated quantum systems}, Nature \textbf{452}, 854 (2008).

\bibitem{Reimann2015} P. Reimann, \textit{Eigenstate thermalization: Deutsch's approach and beyond}, New. J. Phys. \textbf{17}, 055025 (2015).

\bibitem{Reimann2018} P. Reimann, \textit{Dynamical typicality approach to eigenstate thermalization}, Phys. Rev. Lett. {\bf 120}, 230601 (2018).

\bibitem{Deutsch2018} J. M. Deutsch, \textit{Eigenstate thermalization hypothesis}, Rep. Prog. Phys. \textbf{81} 082001 (2018).

\bibitem{erh} T. N. Ikeda, Y. Watanabe, and M. Ueda, \textit{Eigenstate randomization hypothesis: Why does the long-time average equal the microcanonical average?}, Phys. Rev. E \textbf{84}, 021130 (2011).

\bibitem{Altman2018} E. Altman, \textit{ Many-body localization and quantum thermalization}, Nat. Phys. {\bf 14}, 979 (2018).

\bibitem{Basko2006} D. Basko, I. Aleiner, and B. Altshuler, \textit{Metal–insulator transition in a weakly interacting many-electron system with localized single-particle states}, Ann. Phys. \textbf{321}, 1126 (2006).

\bibitem{Alet2018} F. Alet, N. Laflorencie, Many-body localization: An introduction and selected topics, Comp. Rend. Phys. \textbf{19}, 498 (2018).

\bibitem{Abanin2019} D. A. Abanin, J. H. Bardarson, G. D. Tomasi,
S. Gopalakrishnan, V. Khemani, S. A. Parameswaran,
F. Pollmann, A. C. Potter, M. Serbyn, and R. Vasseur, \textit{Distinguishing localization from chaos: challenges in finite-size systems}, arXiv:1911.04501.

\bibitem{Schreiber2015} M. Schreiber, S. S. Hodgman, P. Bordia, H.P. L\"uschen, M.H. Fischer, R. Vosk, E. Altman, U. Schneider, I. Bloch, \textit{Observation of many-body localization of interacting fermions in a quasirandom optical lattice}, Science \textbf{349}, 842 (2015).

\bibitem{Lukin2019} A. Lukin, M.Rispoli, R. Schittko, M. E. Tai, A.M. Kaufman, S. Choi, V. Khemani, J. L\'eonard, 
M. Greiner, \textit{Probing entanglement in a many-body-localized system}, Science \textbf{364}, 256 (2019).

\bibitem{Choi2016} J.-Y. Choi, S. Hild, J. Zeiher, P. Schauß, A. Rubio-Abadal, T. Yefsah, V. Khemani, D. A. Huse, I. Bloch, C. Gross, \textit{Exploring the many-body localization transition in two dimensions}, Science \textbf{352}, 1547 (2016).

\bibitem{Smith2016} J. Smith, A. Lee, P. Richerme, B. Neyenhuis, P. W. Hess, P. Hauke, M. Heyl, D. A. Huse, C. Monroe, \textit{Many-body localization in a quantum simulator with programmable random disorder}, Nat. Phys. \textbf{12}, 907 (2016).

\bibitem{Roushan2017} P. Roushan, C. Neill,  J. Tangpanitanon, V. M. Bastidas, A. Megrant, R. Barends, Y. Chen, Z. Chen, B. Chiaro, A. Dunsworth, A. Fowler, B. Foxen, M. Giustina, E. Jeffrey, J. Kelly, E. Lucero, J. Mutus, M. Neeley, C. Quintana, D. Sank, A. Vainsencher, J. Wenner, T. White, H. Neven, D. G. Angelakis, J. Martinis, \textit{Spectroscopic signatures of localization with interacting photons in superconducting qubits}, Science \textbf{358}, 1175 (2017).

\bibitem{Xu2018} K. Xu, J.-J. Chen, Y. Zeng, Y.-R. Zhang, C. Song, W. Liu, Q. Guo, P. Zhang, D. Xu, H. Deng, K. Huang, H. Wang, X. Zhu, D. Zheng, and H. Fan, \textit{Emulating many-body localization with a superconducting  quantum  processor}, Phys. Rev. Lett. \textbf{120} 050507 (2018).

\bibitem{Suntajs2019} J. \v{S}untajs, J. Bon\v{c}a, T. Prosen, and L. Vidmar, \textit{Quantum chaos challenges many-body localization}, arXiv:1905.06345v2 [cond-mat.str-el].

\bibitem{Sierant2019} P. Sierant, D. Delande, and J. Zakrewski, \textit{Thouless time analysis of Anderson and many-body localization transitions}, arXiv:1911.06221v1 [cond-mat.dis-nn] (2019).

\bibitem{Altshuler1997} B. L. Altshuler, Y. Gefen, A. Kamenev, and L. S. Levitov, \textit{ Quasiparticle Lifetime in a Finite System: A Nonperturbative Approach}, Phys. Rev. Lett. \textbf{78}, 2803 (1997).

\bibitem{Luitz2015} D. J. Luitz, N. Laflorencie, and F. Alet, \textit{ Many-body localization edge in the random-field Heisenberg chain}, Phys. Rev. B \textbf{91}, 081103 (2015).

\bibitem{Facoetti2016} D. Facoetti, P. Vivo, and G. Biroli, \textit{ From non-ergodic eigenvectors to local resolvent statistics and back: A random matrix perspective}, EPL \textbf{115}, 47003 (2016).

\bibitem{Pino2016} M. Pino, L. B. Ioffe, B. L. Altshuler, \textit{ Nonergodic metallic and insulating phases of Josephson junction chains}, Proc. Nat.
Acad. Sci. \textbf{113}, 536 (2016).

\bibitem{Luitz2017} D. J. Luitz and Y. Bar Lev, \textit{The ergodic side of the many-body localization transition}, Ann. Phys. (Berlin) \textbf{529},   1600350 (2017).

\bibitem{Mace2019} N. Mac\'e, F. Alet, and N. Laflorencie, \textit{ Multifractal scalings across the many-body localization transition}, Phys. Rev. Lett. {\bf 123}, 180601 (2019).

\bibitem{Filippone2016} M. Filippone, P. W. Brouwer, J. Eisert, and F. von Oppen, \textit{Drude weight fluctuations in many-body localized systems}, Phys. Rev. B \textbf{94}, 201112(R) (2016).

\bibitem{Sierant2019PRB} P. Sierant and J. Zakrzewski, \textit{Level statistics across the many-body localization transition}, Phys. Rev. B \textbf{99}, 104205 (2019).

\bibitem{Lev2016} Y. B. Lev, D. R. Reichman, and Y. Sagi, \textit{Many-body localization in system with a completely delocalized single-particle spectrum}, Phys. Rev. B \textbf{94}, 201116(R) (2016).

\bibitem{Suntajs2020} J. Šuntajs, J. Bonča, T. Prosen, and L. Vidmar, \textit{Ergodicity Breaking Transition in Finite Disordered Spin Chains}, arXiv:2004.01719 [cond-mat.dis-nn].

\bibitem{Agarwal2015} K. Agarwal, S. Gopalakrishnan, M. Knap, M. M\"{u}ller, E. Demler, \textit {Anomalous Diffusion and Griffiths Effects Near the Many-Body Localization Transition}, Phys. Rev. Lett. \textbf{114}, 160401 (2015).

\bibitem{Gopalakrishnan2016} S. Gopalakrishnan, K. Agarwal, E. A. Demler, D. A. Huse, M. Knap, \textit{ Griffiths effects and slow dynamics in nearly many-body localized systems},
Phys. Rev. B \textbf{93}, 134206 (2016).

\bibitem{Weiner2019} F. Weiner, F. Evers, S. Bera \textit{ Slow dynamics and strong finite-size effects in many-body localization with random and quasiperiodic potentials}, Phys. Rev. B {\bf 100}, 104204 (2019).

\bibitem{Kjall} J. A. Kj\"{a}ll, J. H. Bardarson, and F. Pollmann, Manybody localization in a disordered quantum Ising chain,
Phys. Rev. Lett. 113, 107204 (2014).

\bibitem{Devakul} T. Devakul and R. R. P. Singh, Early breakdown of
area-law entanglement at the many-body delocalization
transition, Phys. Rev. Lett. 115, 187201 (2015).

\bibitem{Luitz2016} D. J. Luitz, Long tail distributions near the many-body
localization transition, Phys. Rev. B 93, 134201 (2016).

\bibitem{Yu2016} X. Yu, D. J. Luitz, and B. K. Clark, Bimodal entanglement entropy distribution in the many-body localization
transition, Phys. Rev. B 94, 184202 (2016).

\bibitem{Modak2015} R. Modak and S. Mukerjee, Many-body localization in
the presence of a single-particle mobility edge, Phys.
Rev. Lett. 115, 230401 (2015).

\bibitem{Serbyn2015} M. Serbyn, Z. Papic, and D. A. Abanin, Criterion for
many-body localization-delocalization phase transition,
Phys. Rev. X 5, 041047 (2015).

\bibitem{Serbyn2017} M. Serbyn, Z. Papic, and D. A. Abanin, Thouless energy
and multifractality across the many-body localization
transition, Phys. Rev. B 96, 104201 (2017).

\bibitem{Kokalj} O. S. Barisic, J. Kokalj, I. Balog, and P. Prelovsek,
Dynamical conductivity and its fluctuations along the
crossover to many-body localization, Phys. Rev. B 94,
7
045126 (2016).

\bibitem{Ros2015} V. Ros, M. M\"{u}ller, and A. Scardicchio, Integrals of motion in the many-body localized phase, Nuc. Phys. B
891, 420 (2015).

\bibitem{Vosk2015} R. Vosk, D. A. Huse, and E. Altman, Theory of the
many-body localization transition in one-dimensional
systems, Phys. Rev. X 5, 031032 (2015).

\bibitem{Potter2015} A. C. Potter, R. Vasseur, and S. A. Parameswaran,
Universal properties of many-body delocalization transitions, Phys. Rev. X 5, 031033 (2015).

\bibitem{Altshuler1986} B. L. Al'tshuler and B. I. Shklovskii, \textit{Repulsion of energy levels and conductivity of small metal samples}, Zh. Eksp. Teor. Fiz. \textbf{91}, 220 (1986) [Sov. Phys JETP \textbf{64}, 127 (1986)].

\bibitem{Altshuler1988} B. L. Al'tshuler, I. Kh. Zharekeshev, S. A. Kotochigova, and B. I. Shklovskii \textit{Repulsion between energy levels and the metal insulator transition}, Zh. Eksp. Teor. Fiz. \textbf{94}, 343 (1988) [Sov. Phys JETP \textbf{67}, 625 (1988)].

\bibitem{stockmann} H. J. St\"{o}ckmann,
\textit{Quantum chaos}
(Cambridge University Press, Cambridge, 1999).

\bibitem{Jensen1985} R. V. Jensen and R. Shankar, \textit{Statistical behavior in deterministic quantum systems with few degrees of freedom}, Phys. Rev. Lett. \textbf{54}, 1879 (1985).

\bibitem{Srednicki1994} M. Srednicki, \textit{Chaos and quantum thermalization}, Phys. Rev. E \textbf{50}, 888 (1994).

\bibitem{Deutsch1991} J. M. Deutsch, \textit{Quantum statistical mechanics in a closed system}, Phys. Rev. A \textbf{43}, 2046 (1991).

\bibitem{Edwards1972} J. T. Edwards, D. J. Thouless, \textit{ Numerical studies of localization in disordered systems}, J. Phys. C: Solid State Phys. {\bf 5}, 807 (1972).

\bibitem{Shapiro1993} B. I. Shklovskii, B. Shapiro, B. R. Sears, P. Lambrianides, and H. B. Shore, \textit{Statistics of spectra of disordered systems near the metal-insulator transition}, Phys. Rev. B \textbf{47}, 11487 (1993).

\bibitem{Griffiths1969} R. B. Griffiths, \textit{Nonanalytic behavior above the critical
point in a random ising ferromagnet}, Phys. Rev. Lett.
\textbf{23}, 17 (1969).

\bibitem{Luschen2017} H. P. Lüschen, P. Bordia, S. Scherg, F. Alet, E. Altman, U. Schneider, and I. Bloch, \textit{Observation of Slow Dynamics near the Many-Body Localization Transition in One-Dimensional Quasiperiodic Systems}, Phys. Rev. Lett. \textbf{119}, 260401 (2017).

\bibitem{Reimann2008} P. Reimann, \textit{Foundations of statistical mechanics under experimentally realistic conditions}, Phys. Rev. Lett. \textbf{101}, 190403 (2008).

\bibitem{Linden2009} N. Linden, S. Popescu, A. J. Short, and A. Winter, \textit{Quantum mechanical evolution towards thermal equilibrium}, Phys. Rev. E \textbf{79}, 061103 (2009).

\bibitem{Hamazaki2018} R. Hamazaki and M. Ueda, \textit{Atypicality of most few-body observables}, Phys. Rev. Lett. {\bf 120}, 080603 (2018).

\bibitem{Pathria} R. K. Pathria and P. D. Beale, \textit{Statistical Mechanics} (Elsevier, 3rd Edition, 2011).

\bibitem{Peres1984} A. Peres, \textit{New conserved quantities and test for regular spectra}, Phys. Rev. Lett. \textbf{53}, 1711 (1984).

\bibitem{Lobez2016} C. M. L\'obez and A. Rela\~{n}o, \textit{Entropy, chaos, and excited-state quantum phase transitions in the Dicke model}, Phys. Rev. E {\bf 94}, 012140 (2016).

\bibitem{Bertrand2016} C. L. Bertrand and A. M. Garc\'{i}a-Garc\'{i}a, \textit{Anomalous Thouless energy and critical statistics on the metallic side of the many-body localization transition}, Phys. Rev. B \textbf{94}, 144201 (2016).

\bibitem{Serbyn16} M. Serbyn and J. E. Moore, \textit{Spectral statistics across the many-body localization transition}, Phys. Rev. B {\bf 93}, 041424 (2016).

\bibitem{Torres2017} E. J. Torres-Herrera and L. F. Santos, \textit{Extended nonergodic states in disordered many-body quantum systems}, Ann. Phys. (Berlin) \textbf{529}, No. 7, 1600284 (2017).

\bibitem{lfsantos} L. F. Santos, \textit{Integrability of a disordered Heisenberg spin-$1/2$ chain}, J. Phys. A: Math. Gen. \textbf{37} (2004) 4723.

\bibitem{Buijsman2019} W. Buijsman, V. Cheianov, and V. Gritsev, \textit{Random matrix ensemble for the level statistics of many-body localization}, Phys. Rev. Lett. \textbf{122}, 180601 (2019).

\bibitem{Lev2014} Y. Bar Lev and D. R. Reichman, Phys.Rev. B \textbf{89}, 220201 (2014).

\bibitem{Lev2015} Y. Bar Lev, G. Cohen, and D. R. Reichman, Phys. Rev. Lett. \textbf{114}, 100601 (2015).

\bibitem{Znidaric2016} M. Znidaric, A. Scardicchio, and V. K. Varma, Phys. Rev. Lett. \textbf{117}, 040601 (2016)

\newpage{\pagestyle{empty}\cleardoublepage}

\bibitem{ratios} Y. Y. Atas, E. Bogomolny, O. Giraud, and G. Roux, \textit{Distribution of the ratio of consecutive level spacings in random matrix ensembles}, Phys. Rev. Lett. \textbf{110}, 084101 (2013).

\bibitem{Oganesyan2007} V. Oganesyan and D. A. Huse, Localization of interacting
fermions at high temperature, Phys. Rev. \textbf{B} 75, 155111 (2007)

\bibitem{Corps2020} A. L. Corps and A. Rela\~{n}o, \textit{Distribution of the ratio of consecutive level spacings for different symmetries and degrees of chaos}, Phys. Rev. E \textbf{101}, 022222 (2020).

\bibitem{conjetura} A. Rela\~{n}o, J. M. G. G\'{o}mez, R. A. Molina, J. Retamosa, and E. Faleiro, \textit{Quantum chaos and $1/f$ noise}, Phys. Rev. Lett. \textbf{89}, 244102 (2002).

\bibitem{demo} E. Faleiro, J. M. G. G\'{o}mez, R. A. Molina, L. Mu\~{n}oz, A. Rela\~{n}o and J. Retamosa, \textit{Theoretical derivation of $1/f$ noise in quantum chaos}, Phys. Rev. Lett. \textbf{93}, 244101 (2004).

\bibitem{misleadingsign} J. M. G. G\'{o}mez, R.A. Molina, A. Rela\~{n}o, and J. Retamosa, \textit{Misleading signatures of quantum chaos}, Phys. Rev. E \textbf{66}, 036209 (2002).

\bibitem{Pachon2018} L. A. Pachon, A. Rela\~{n}o, B. Peropadre, A. Aspuru-Guzik, \textit{Origin of the $1/f^{\alpha}$-Spectral-Noise in Chaotic and Regular Quantum Systems}, Phys. Rev. E 98, 042213 (2018).

\bibitem{Gomez2005} J. M. G. G\'{o}mez, A. Rela\~{n}o, J. Retamosa, E. Faleiro, L. Salasnich, M. Vranicar, and M. Robnik, \textit{$1/f^{\alpha}$ Noise in Spectral Fluctuations of Quantum Systems}, Phys. Rev. Lett. \textbf{94}, 084101 (2005).

\bibitem{Santhanam2005} M. S. Santhanam and J. N. Bandyopadhyay, \textit{Spectral Fluctuations and $1/f$ Noise in the Order-Chaos Transition Regime}, Phys. Rev. Lett. {\bf 95}, 114101 (2005).

\bibitem{Relano2008} A. Rela\~{no}, \textit{Chaos-Assisted Tunneling and $1/f^{\alpha}$ Spectral Fluctuations in the Order-Chaos Transition}, Phys. Rev. Lett. {\bf 100}, 224101 (2008).

\bibitem{Relano2008b} A. Rela\~{n}o, L. Mu\~{n}oz, J. Retamosa, E. Faleiro, and R. A. Molina, \textit{Power-spectrum characterization of the continuous Gaussian ensemble}, Phys. Rev. E {\bf 77}, 031103 (2008).

\bibitem{Schiulaz2019} M. Schiulaz, E. J. Torres-Herrera, and L. F. Santos, \textit{Thouless and relaxation time scales in many-body quantum systems}, Phys. Rev. B \textbf{99}, 174313 (2019).

\bibitem{Santos2010} L. F. Santos and M. Rigol, \textit{Onset of quantum chaos in one-dimensional bosonic and fermionic systems and its relation to thermalization}, Phys. Rev. E \textbf{81}, 036206 (2010).

\bibitem{Torres-Herrera2014} E. J. Torres-Herrera and L. F. Santos, \textit{Local quenches with global effects in interacting quantum systems}, Phys. Rev. E \textbf{89}, 062110
(2014).

\bibitem{shiftinvert} F. Pietracaprina, N. Macé, D. J. Luitz, and F. Alet,
\textit{Shift-invert diagonalization of large many-body localizing spin chains}, SciPost Phys. \textbf{5}, 045 (2018).


\end{thebibliography}
\end{document}